\documentclass{article}

\PassOptionsToPackage{table,xcdraw}{xcolor}

\usepackage{microtype}
\usepackage{graphicx}
\usepackage{subcaption}
\usepackage{booktabs} 

\usepackage[
    colorlinks=true,
    linkcolor=black,
    urlcolor=mypink,
    citecolor=cyan
]{hyperref}

\let\evoharnessaddcontentsline\addcontentsline
\usepackage{salesforce}
\let\addcontentsline\evoharnessaddcontentsline

\usepackage{amsmath}
\usepackage{amssymb}
\usepackage{mathtools}
\usepackage{amsthm}

\theoremstyle{plain}

\theoremstyle{definition}

\theoremstyle{remark}

\usepackage[utf8]{inputenc} 
\usepackage[T1]{fontenc}    
\usepackage{url}            
\usepackage{booktabs}       
\usepackage{amsfonts}       
\usepackage{nicefrac}       
\usepackage{microtype}      

\usepackage{inconsolata}
\usepackage{url}
\usepackage{booktabs}       
\usepackage{amsfonts}       
\usepackage{nicefrac}       
\usepackage{microtype}      
\usepackage{url}            
\usepackage{booktabs}       
\usepackage{amsfonts}       
\usepackage{nicefrac}       
\usepackage{microtype}      
\usepackage{subcaption}
\usepackage{amsmath,amsfonts,amssymb}
\usepackage{tabularx}
\usepackage{multirow}
\usepackage{graphicx}
\usepackage{color}
\usepackage{CJK}
\usepackage{adjustbox}
\usepackage{colortbl}
\usepackage{multicol}
\usepackage{vwcol} 
\newsavebox{\algleft}
\newsavebox{\algright}
\usepackage{bm}
\usepackage{booktabs}
\usepackage{amsmath,stackengine}
\usepackage{times}
\usepackage{latexsym}
\usepackage{bbm}
\usepackage[T1]{fontenc}    
\usepackage{url}            
\usepackage{booktabs}       
\usepackage{amsfonts}       
\usepackage{nicefrac}       
\usepackage{microtype}      
\usepackage{amssymb}
\usepackage{pifont}
\usepackage{bbm}
\usepackage{breqn}
\usepackage{longtable}
\usepackage{enumitem}
\usepackage{xcolor}
\usepackage{microtype}
\usepackage{enumitem}
\usepackage{amsmath}
\usepackage{chngcntr} 
\usepackage{wrapfig}
\usepackage{listings}
\usepackage{arydshln}
\usepackage{amsmath} 

\usepackage{booktabs}
\usepackage{longtable}
\usepackage{xspace}

\usepackage{lipsum}

\usepackage[most]{tcolorbox}
\usepackage{makecell}
\usepackage{titletoc}

\usepackage[capitalize,noabbrev]{cleveref}

\usepackage{fontawesome5}

\usepackage[textwidth=2.5cm]{todonotes}

\usepackage{makecell}

\definecolor{BlueBG}{rgb}{0,0.46,0.71}

\newcommand{\shafiq}[1]{{\color{red}{\small\bf\sf [shafiq: #1]}}}

\newcommand{\ourbenchmark}{{\textsc{EvoHarnessBench}}\xspace}

\definecolor{innerboxcolor}{HTML}{00B2E6} 
\definecolor{outerboxcolor}{HTML}{D7F6FF}
\definecolor{lightblue}{HTML}{D7F6FF}
\definecolor{mypink}{RGB}{190, 50, 110}

\usepackage{tikz}
\usetikzlibrary{decorations.pathmorphing}

\definecolor{upblue}{HTML}{388E3C}
\definecolor{downred}{HTML}{E53935}

\definecolor{darkgreen}{RGB}{0,128,0}  
\definecolor{darkred}{RGB}{139,0,0}

\definecolor{myblue}{RGB}{31,119,180}   
\definecolor{mygreen}{RGB}{44,160,44}    
\definecolor{myorange}{RGB}{255,127,14}
\definecolor{mypurple}{RGB}{153, 102, 204}

\newtcolorbox[auto counter]{observation}[1][]{
  breakable,
  colback=black!5!white,
  colframe=black!70!white,
  fonttitle=\bfseries,
  enhanced,
  boxrule=0.6pt,
  left=1mm,right=1mm,top=1mm,bottom=1mm,
  #1
}

\newtcolorbox[auto counter]{takeaway}[1][]{
  breakable,
  colback=teal!3!white,
  colframe=teal!55!black,
  fonttitle=\bfseries,
  title=Takeaway~\thetcbcounter,
  enhanced,
  boxrule=0.5pt,
  left=1mm,right=1mm,top=1mm,bottom=1mm,
  #1
}

\newtcolorbox[auto counter]{practicalguidance}[1][]{
  breakable,
  colback=cyan!3!white,
  colframe=cyan!60!black,
  fonttitle=\bfseries,
  title=Practical Guidance~\thetcbcounter,
  enhanced,
  boxrule=0.5pt,
  left=1mm,right=1mm,top=1mm,bottom=1mm,
  #1
}

\newtcolorbox[auto counter]{discussion}[1][]{
  breakable,
  colback=violet!4!white,
  colframe=violet!60!black,
  fonttitle=\bfseries,
  title=Discussion~\thetcbcounter,
  enhanced,
  boxrule=0.5pt,
  left=1mm,right=1mm,top=1mm,bottom=1mm,
  #1
}

\newtcolorbox[auto counter]{developprompt}[1][]{
  breakable,
    colback=white,
    colframe=black!70,
  fonttitle=\bfseries,
  title=Prompt~\thetcbcounter,
  enhanced,
  boxrule=0.5pt,
  left=1mm,right=1mm,top=1mm,bottom=1mm,
  #1
}

\newtcolorbox[auto counter]{systemprompt}[1][]{
  breakable,
    colback=orange!6,
    colframe=orange!60!black,
  fonttitle=\bfseries,
  title=System Prompt~\thetcbcounter,
  enhanced,
  boxrule=0.5pt,
  left=1mm,right=1mm,top=1mm,bottom=1mm,
  #1
}

\newtcolorbox[auto counter]{userprompt}[1][]{
  breakable,
    colback=green!5,
    colframe=green!40!black,
  fonttitle=\bfseries,
  title=User Prompt~\thetcbcounter,
  enhanced,
  boxrule=0.5pt,
  left=1mm,right=1mm,top=1mm,bottom=1mm,
  #1
}

\newtcolorbox[auto counter]{answerprompt}[1][]{
  breakable,
  colback=purple!6,
  colframe=purple!50!black,
  fonttitle=\bfseries,
  title=Answer Prompt~\thetcbcounter,
  enhanced,
  boxrule=0.5pt,
  left=1mm,right=1mm,top=1mm,bottom=1mm,
  #1
}
\newtcolorbox[auto counter]{subagent}[1][]{
  breakable,
  colback=cyan!3!white,
  colframe=cyan!60!black,
  fonttitle=\bfseries,
  title=Sub-agent~\thetcbcounter,
  enhanced,
  boxrule=0.5pt,
  left=1mm,right=1mm,top=1mm,bottom=1mm,
  #1
}

\newcolumntype{Y}{>{\centering\arraybackslash}X}

\newcommand{\projectpage}{\url{https://mas-orchestra.salesforceresearch.ai/evoharness/}}

\definecolor{agentrow}{HTML}{F0F6EE}
\definecolor{modelrow}{HTML}{FFF6E8}
\definecolor{harnessrow}{HTML}{EEF5FB}
\definecolor{termrow}{HTML}{F5F0FA}
\definecolor{oracletrow}{HTML}{E6E6E6}
\definecolor{lastexamred}{HTML}{D00000}
\definecolor{sc0}{HTML}{9C2F2F}
\definecolor{sc1}{HTML}{B45F06}
\definecolor{sc2}{HTML}{8A6D1D}
\definecolor{sc3}{HTML}{4F7C2C}
\definecolor{sc4}{HTML}{2E6F3E}
\definecolor{sc5}{HTML}{1D5B35}
\newcommand{\hsc}[1]{%
  \ifdim #1 pt < 5pt \textcolor{sc0}{#1}%
  \else\ifdim #1 pt < 20pt \textcolor{sc1}{#1}%
  \else\ifdim #1 pt < 40pt \textcolor{sc2}{#1}%
  \else\ifdim #1 pt < 60pt \textcolor{sc3}{#1}%
  \else\ifdim #1 pt < 80pt \textcolor{sc4}{#1}%
  \else \textcolor{sc5}{#1}%
  \fi\fi\fi\fi\fi}
\newcommand{\hscb}[1]{%
  \ifdim #1 pt < 5pt \textcolor{sc0}{\textbf{#1}}%
  \else\ifdim #1 pt < 20pt \textcolor{sc1}{\textbf{#1}}%
  \else\ifdim #1 pt < 40pt \textcolor{sc2}{\textbf{#1}}%
  \else\ifdim #1 pt < 60pt \textcolor{sc3}{\textbf{#1}}%
  \else\ifdim #1 pt < 80pt \textcolor{sc4}{\textbf{#1}}%
  \else \textcolor{sc5}{\textbf{#1}}%
  \fi\fi\fi\fi\fi}

\definecolor{papertodobg}{HTML}{FFF8E1}
\definecolor{papertodoframe}{HTML}{E65100}

\newtcolorbox{lastexambox}{%
  enhanced, sharp corners,
  colback=lastexamred!4, colframe=lastexamred!70!black,
  boxrule=0pt, leftrule=3pt,
  left=7pt, right=7pt, top=5pt, bottom=5pt,
  before skip=6pt, after skip=8pt,
  fontupper=\small
}

\graphicspath{{./}}
\definecolor{halinkcol}{HTML}{1B4F9C}

\begin{document}

\title{\ourbenchmark: \\ Can Your Agents Keep Pace with an Evolving Harness?}

\author{ 
Zixuan Ke\thanks{ 
Equal contribution. 
$^{\dagger}$Core contributors. 
$^{\ddagger}$Senior authors. 
$^1$Salesforce Research. 
$^2$University of North Carolina at Chapel Hill, conducted this work during an internship at Salesforce Research.
$^3$University of North Carolina at Chapel Hill
$^4$University of Wisconsin--Madison.}$^{*\dagger 1}$
~~~ Vaidehi Patil$^{*\dagger 2}$
~~~ Haizhou Shi$^{*\dagger 1}$
~~~ Yang Li$^{\dagger 1}$
~~~ Ye Liu$^{1}$
\AND Sarath Shekkizhar$^{1}$
~~~ Anurag Koul$^{1}$
~~~ Jiayu Wang$^{4}$
~~~ Xuan Phi Nguyen$^{1}$
\AND Semih Yavuz$^{\ddagger 1}$
~~~ Mohit Bansal$^{\ddagger 3}$
~~~ Shafiq Joty$^{\ddagger 1}$  }

\maketitle
\vspace{-2\baselineskip}

\begin{center}
    \textcolor{mypink}{\textbf{\faIcon{link} \small \projectpage}}
\end{center}


\begin{abstract}

Modern LLM-based agents operate through a \emph{harness} of tools, reusable skills, and specialist agents that shapes what they observe and what they can do. In practice, this harness is inherently non-stationary, continually expanding as new capabilities are introduced. But can agents keep pace with an evolving harness? We introduce \textbf{\ourbenchmark}, a benchmark for systematically evaluating agents under controlled harness evolution along three axes: \textit{tools}, \textit{skills}, and \textit{agents}. Unlike existing continual-learning benchmarks for agents, which typically place non-stationarity in the task stream while keeping the harness fixed, \ourbenchmark places non-stationarity in the externally supplied harness itself. The benchmark contains 17 multi-stage harness streams constructed deterministically from verifier-based benchmarks, comprising \textbf{802 tasks, 520 tools, 42 skills, and 62 agents}. We evaluate two complementary settings that capture the central challenges of harness evolution: \emph{deployment evaluation}, which measures whether agents retain previously accessible competence as the harness expands, and \emph{self-evolving adaptation evaluation}, which tests whether accumulated experience remains useful as new capabilities are introduced. Our experiments reveal three persistent limitations in state-of-the-art agents. First, harness expansion alone can degrade performance on previously solved tasks, leading to \emph{harness-induced forgetting}. Second, gains from self-evolving adaptation are inconsistent across stages, capability axes, and environments. Third, retention and adaptation can be in tension: preserving earlier competence does not necessarily improve adaptation to newly introduced capabilities, and vice versa. Together, these results establish harness evolution as a distinct and important challenge for building agents that can continuously exploit new capabilities while preserving previously effective behavior.

\end{abstract}

\section{Introduction}
\label{sec.intro}

\begin{figure}[t]
\centering
\includegraphics[width=\columnwidth]{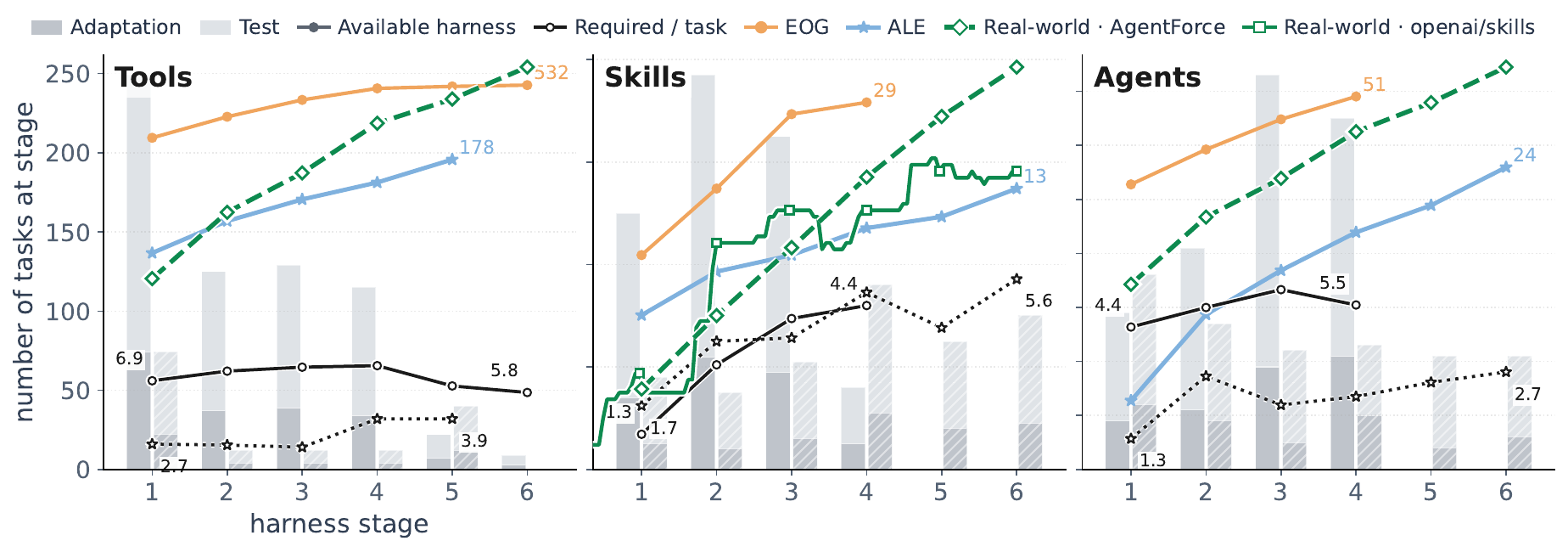}
\vspace{-2\baselineskip}
\captionof{figure}{\small Statistics of \ourbenchmark. Green curves are real-world references ordered by capability release date. \ourbenchmark follows the same qualitative pattern: the available harness grows substantially over time while individual tasks require only a small subset of it. 
}
\label{fig:dataset}
\vspace{-1\baselineskip}
\end{figure}

LLM-based agentic systems involve much more than a model and a prompt. They operate through a \textbf{harness}: the tools they can invoke, the skills or procedures they can reuse, and the other agents they can coordinate with~\citep{gu2026modelscalingscalingscaling,zhou2026externalizationllmagentsunified,macedo2026stophandholdingcodingagent,ke2025surveyfrontiersllmreasoning}. This harness fundamentally shapes both what an agent \emph{observes} and what it can \emph{do}. Importantly, the harness presented to an agent is not static. In deployed systems, it is inherently \textbf{non-stationary}, continually changing as new tools, skills, and specialist agents are introduced, updated, or retired. For example, the Agentforce ecosystem of Salesforce has expanded steadily\footnote{\url{https://agentexchange.salesforce.com/explore/agentforce}} (\cref{fig:dataset}, green dashed line), while the history of OpenAI's public skills repository reflects a continuing stream of capability additions, updates, and removals\footnote{\url{https://github.com/openai/skills}} (\cref{fig:dataset}, green line). These trends suggest that harness evolution is not a peripheral edge case, but an increasingly central property of real-world agentic systems.

\begin{figure}[t]
\centering
\includegraphics[width=0.9\columnwidth]{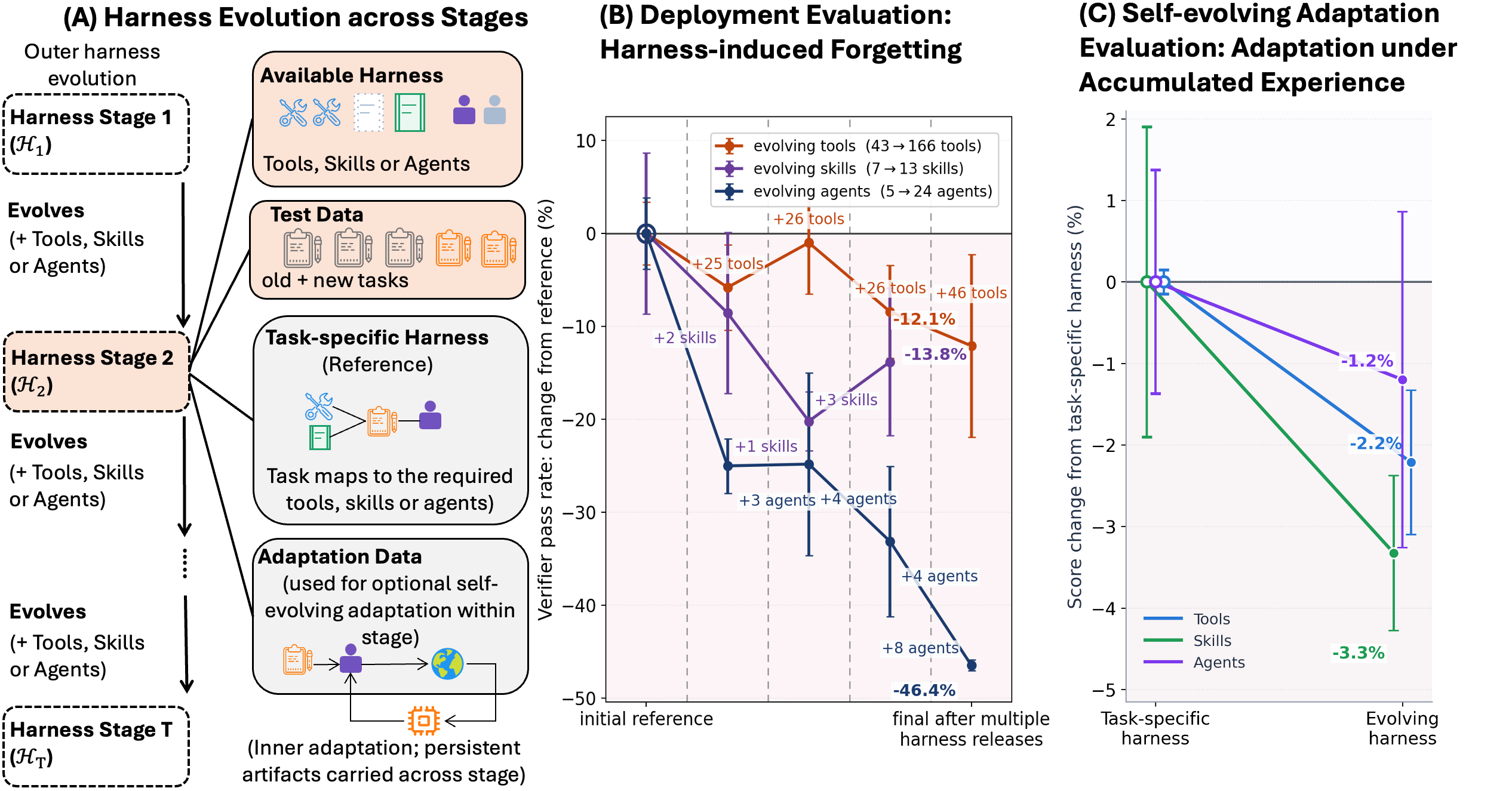}
\captionof{figure}{
\small \textbf{(A)} \ourbenchmark evaluates agents under \emph{outer harness evolution}, where the externally supplied harness expands across stages ($\mathcal{H}_1,\mathcal{H}_2,\ldots,\mathcal{H}_T$) along one capability axis: tools, skills, or specialist agents. At each stage, the \emph{available harness} contains the full cumulative capability pool, while the \emph{test data} include both newly introduced and previously encountered task cohorts. A \emph{task-specific harness}, containing only the capabilities required for each task, serves as a reference condition. An \emph{adaptation split} is used optionally for inner self-evolving adaptation, with persistent artifacts carried across stages. This yields two settings: \emph{deployment evaluation}, without inner adaptation, and \emph{self-evolving adaptation evaluation}, with it. \textbf{(B)} Deployment evaluation reveals \emph{harness-induced forgetting} caused by harness evolution alone, with performance drops of $12.1\%$, $13.8\%$, and $46.4\%$ for evolving tools, skills, and agents, respectively. \textbf{(C)} Self-evolving adaptation evaluation shows that adapting under an evolving harness can be more challenging than under task-specific reference harnesses, with the largest degradation for skills ($-3.3\%$), followed by tools ($-2.2\%$) and agents ($-1.2\%$). 
}
\label{fig:overview}
\vspace{-1\baselineskip}
\end{figure}

Harness evolution creates two distinct challenges.
 (1) \textbf{Retention}: as the externally supplied harness expands, an agent must continue to solve tasks that were previously supported. The capabilities required for those tasks may still be present, but they are now embedded within a larger pool of tools, skills, or agents, making previously successful behavior harder to recover and execute reliably. This can lead to \textbf{harness-induced forgetting}: the underlying model parameters remain unchanged, yet previously accessible competence degrades solely because the harness through which the agent acts has expanded (\cref{fig:overview},(B)). (2) \textbf{Adaptation under accumulated experience}: in systems with \emph{self-evolving adaptation}\footnote{We focus on \textit{harness-based} self-evolving adaptation: the underlying model weights remain fixed, while persistent components surrounding the model may be updated.}, persistent artifacts such as memories, learned skills, prompts, or routing policies are accumulated under \textbf{earlier}, \textbf{narrower} harnesses. As new capabilities are introduced, these artifacts may become stale or even actively misdirect execution toward behaviors that were effective under previous harnesses but are no longer appropriate under the current one (\cref{fig:overview},(C)).

To expose and systematically evaluate these two challenges, we introduce \textbf{\ourbenchmark}, a benchmark specifically designed to study \textbf{harness evolution}. To our knowledge, \ourbenchmark is the \textbf{first benchmark} to evaluate how growth in an externally supplied harness affects agent performance over time. Unlike prior benchmarks~\citep{li2026skillsbenchbenchmarkingagentskills,zhong2026skilllearnbenchbenchmarkingcontinuallearning,asawa2026continuallearningbenchevaluating}, which typically keep the harness fixed, place non-stationarity in the task stream, or study capability growth produced by the agent itself, \ourbenchmark makes the externally supplied harness the source of non-stationarity and measures how agent performance changes as that harness evolves. It constructs nested, cumulative harness streams along three axes: \textit{tools} (expanding API catalogs), \textit{skills} (growing libraries of reusable procedures), and \textit{agents} (expanding pools of specialist agents). Capabilities are introduced progressively from frequently used core capabilities to the long tail, and each task is assigned to the earliest stage at which all of its required capabilities are available, while requiring at least one capability newly introduced at that stage.

In \ourbenchmark, we adopt two complementary evaluation modes, \textbf{deployment evaluation} and \textbf{self-evolving adaptation evaluation}, corresponding to the two challenges posed by harness evolution. Both are organized around \textbf{outer harness evolution}, where the externally supplied harness expands across stages  as new tools, skills, or agents are introduced, with optional \textbf{inner self-evolving adaptation} within each stage (\cref{fig:overview}A). In \textit{deployment evaluation}, inner adaptation is disabled: each harness version is evaluated independently, with no experience carried across stages. This setting isolates the direct effect of harness evolution, testing whether frontier agents can exploit newly available capabilities while remaining robust to the growing pool of alternatives and potential distractions. In \textit{self-evolving adaptation evaluation}, inner adaptation is enabled: at each stage, the system may learn through a self-evolving adaptation method and carry the resulting persistent artifacts, such as memories, learned skills, or prompts, across subsequent harness versions, while evaluation is performed on a held-out split. This setting measures whether experience accumulated under earlier harnesses continues to support effective adaptation as new capabilities are introduced.

In total, \ourbenchmark contains \textit{17 controlled harness streams}, each comprising \textit{3 to 6 cumulative stages}. It includes \textit{802 unique tasks}, instantiated as \textit{1,510 axis-specific evaluation examples} spanning \textit{520 executable tools}, \textit{42 latent reference skills}, and \textit{62 specialist agents}. We evaluate representative single-agent systems (SAS) and multi-agent systems (MAS) under both evaluation modes. Our experiments reveal three persistent gaps in state-of-the-art agents. (1) \textbf{Harness evolution is not free}: frontier agents can lose competence on previously solved tasks simply because the externally supplied harness has expanded, even when the capabilities required to solve those tasks remain available.(2)~\textbf{Self-evolving adaptation remains inconsistent}: gains vary substantially across harness stages, capability axes, and environments, indicating that experience accumulated under earlier harnesses does not reliably transfer as the harness evolves. (3)~\textbf{Retention and adaptation can be in tension}: methods that best preserve previously accessible competence may hinder adaptation to newly introduced capabilities, while methods that adapt effectively may sacrifice retention. Together, these findings show that current agents do not yet keep pace reliably with harness evolution. They motivate methods that explicitly couple \emph{outer harness evolution} with \emph{inner adaptation}, rather than optimizing adaptation in isolation, so that agents can preserve earlier competence while effectively exploiting newly introduced capabilities.

\section{Related Work}
\label{sec.related_work}
In this section, we provide a brief overview of the most relevant prior work, with an extended discussion in Appendix~\ref{sec:relate_detail}. As summarized in Table~\ref{tab:related_work}, existing agent benchmarks can be broadly grouped into three categories, none of which directly evaluates how agent performance changes as the harness itself evolves:

\begin{wraptable}{r}{0.63\columnwidth}
\vspace{-1.5\baselineskip}
\caption{\small Comparison with related agent benchmarks.}
\vspace{-1.0em}
\centering
\scriptsize
\setlength{\tabcolsep}{2.5pt}
\renewcommand{\arraystretch}{0.92}
\resizebox{\linewidth}{!}{
\begin{tabular}{lcccc}
\toprule
\textbf{Benchmark}
& \makecell{\textbf{Outer}\\\textbf{Evol.}}
& \makecell{\textbf{Inner}\\\textbf{Adapt.}}
& \makecell{\textbf{Primary}\\\textbf{Evaluation}}
& \textbf{Axes} \\
\midrule

\rowcolor{gray!10}
\multicolumn{5}{l}{\textbf{Static Harness Benchmarks}} \\
API-Bank / $\tau$-bench
& None & No & Task performance & Tools \\
SkillsBench / SkillCraft
& None & No & Task performance & Skills \\
MultiAgentBench
& None & No & Task performance & Agents \\

\midrule
\rowcolor{gray!10}
\multicolumn{5}{l}{\textbf{Benchmarks for Self-Evolving Agents}} \\
Paper-/ RE-/ MLE-Bench
& None & Yes & Task performance & -- \\
SkillLearnBench / SkillFlow
& None & Yes & Learned-skill quality & Skills \\

\midrule
\rowcolor{gray!10}
\multicolumn{5}{l}{\textbf{Continual Agent Benchmarks}} \\
AgentCL / ContinualBench
& Task stream & Yes & Continual adaptation & -- \\
CMRBench
& Model pool & Yes & Continual routing & Models \\

\midrule
\textbf{\ourbenchmark}
& \textbf{Harness}
& \textbf{Optional}
& \makecell{\textbf{Deployment /}\\\textbf{Self-evolving adaptation}}
& \textbf{Tools, Skills, Agents} \\
\bottomrule
\end{tabular}
}
\label{tab:related_work}
\vspace{-1.0em}
\end{wraptable}

(1) \textbf{Static-harness benchmarks} such as $\tau$-bench, SkillsBench and MultiAgentBench evaluate agents with a fixed set of tools, skills, or agents~\citep{
li2023apibankcomprehensivebenchmarktoolaugmented,
yao2024taubenchbenchmarktoolagentuserinteraction,
li2026skillsbenchbenchmarkingagentskills,
zhu2025multiagentbenchevaluatingcollaborationcompetition}; they involve neither outer harness evolution nor inner adaptation.
(2) \textbf{Self-evolving benchmarks} study whether agents improve through accumulated experience~\citep{ouyang2026reasoningbank}. These settings permit inner adaptation but do not include outer harness evolution, and therefore do not capture how accumulated experience behaves as the externally supplied harness changes. Skill-learning benchmarks such as SkillLearnBench~\citep{zhong2026skilllearnbenchbenchmarkingcontinuallearning} and SkillFlow~\citep{zhang2026skillflowbenchmarkinglifelongskilldiscovery} do involve an expanding skill surface, but that growth is agent-driven as part of the adaptation process itself; their primary focus is learned-skill quality rather than task performance under an externally evolving harness. (3) \textbf{Continual-learning benchmarks} include both outer evolution and inner adaptation, but place the source of non-stationarity in the task stream rather than in the harness~\citep{shu2026agentclrigorousevaluationcontinual,
asawa2026continuallearningbenchevaluating}. The closest setting is CMRBench~\citep{bell2026continual}, which studies continual routing over an externally expanding model pool. In contrast, \ourbenchmark places outer evolution directly in the \textit{agent harness}, across tools, skills, and specialist agents. 

\section{\ourbenchmark}
\label{sec:benchmark}

\subsection{Overview}
\label{sec:benchmark_overview}

We formalize harness evolution as follows. Let $
\mathcal{H}_{1:T} \triangleq (\mathcal{H}_1,\mathcal{H}_2,\ldots,\mathcal{H}_T)
$ denote a sequence of harness states satisfying $
\mathcal{H}_1 \subseteq \mathcal{H}_2 \subseteq \cdots \subseteq \mathcal{H}_T,
$, where $\mathcal{H}_t$ is the set of capabilities exposed to the agent at stage $t$. Depending on the evolution axis, each capability may correspond to an executable tool, a procedural skill, or a specialist agent. The sequence $\mathcal{H}_{1:T}$ defines the \emph{outer harness evolution}: capabilities may be added between stages, but the harness remains fixed during execution within a stage.  Associated with each stage $t$ is a task set $
\mathcal{D}_t = \{(x_i,y_i,\mathcal{C}_i)\}_{i \in I_t},
$, where $x_i$ denotes the user task, $y_i$ is its verifier-checkable target outcome,  and $\mathcal{C}_i$ is a hidden ground-truth set of required capabilities used only for benchmark construction and analysis. Once introduced, the task instance $(x_i,y_i)$ remains unchanged at subsequent stages; only the surrounding harness expands. Let $t_i$ denote the stage at which task $i$ is introduced. By construction, every task satisfies:
\begin{enumerate}[leftmargin=*,noitemsep,topsep=0pt]
\item \textbf{Feasibility:} all capabilities needed to solve the task are available at introduction, i.e., $\mathcal{C}_i \subseteq \mathcal{H}_{t_i}$.
\item \textbf{New-capability pressure:} at least one required capability is newly introduced at that stage, i.e.,
$\mathcal{C}_i \cap (\mathcal{H}_{t_i} \setminus \mathcal{H}_{t_i-1}) \neq \emptyset$.
\end{enumerate}

At each stage $t$, the corresponding task set $\mathcal{D}_t$ is partitioned into an adaptation split $\mathcal{D}_{t,\mathrm{adapt}}$ and a held-out evaluation split $\mathcal{D}_{t,\mathrm{eval}}$. Systems with self-evolving adaptation may use $\mathcal{D}_{t,\mathrm{adapt}}$ for \emph{inner adaptation}, whereas $\mathcal{D}_{t,\mathrm{eval}}$ is reserved for verifier-based measurement. During evaluation, the agent receives the full cumulative harness $\mathcal{H}_t$, rather than an oracle-pruned task-specific subset, so previously introduced capabilities remain available as either useful affordances or potential distractors.

We instantiate \ourbenchmark from EnterpriseOps-Gym (\textbf{EOG})~\citep{malay2026enterpriseopsgymenvironmentsevaluationsstateful} and Agents' Last Exam (\textbf{ALE})~\citep{sun2026agentslastexam}. EOG contributes stateful enterprise workflows with structured tool annotations, while ALE contributes diverse agentic tasks with rich software-tool environments. Together, they yield \textbf{17 harness streams} with 3--6 stages each, covering \textbf{802 unique tasks} instantiated as \textbf{1,510 axis-specific evaluation examples} across \textbf{520 tools}, \textbf{42 latent reference skills}, and \textbf{62 specialist agents}. Full benchmark statistics are provided in \cref{sec:statistics}.

\subsection{Construction Pipeline}
\label{sec:construction}

\ourbenchmark converts static agent benchmarks into evolving harness streams without introducing new tasks. The construction pipeline is applied independently to each evolution axis $a \in \{\mathrm{tool}, \mathrm{skill}, \mathrm{agent}\}$; for clarity, we omit the axis superscript below. \cref{fig:construction_pipeline} provides an overview of this  construction process.

\textbf{Step 1: Capability annotation.} \quad
Each task $i$ is associated with an axis-specific capability set $\mathcal{C}_i$. For the tool axis, these annotations are inherited directly from the oracle tool labels provided by the source benchmarks. For the skill and agent axes, where such annotations are not available, we construct them deterministically following the procedures described in \cref{sec:axes}.

\textbf{Step 2: Frequency-ranked capability release.} \quad
We define the capability universe as $\mathcal{U} = \bigcup_i \mathcal{C}_i$ and measure the frequency of each capability $c \in \mathcal{U}$ as
$f(c) = |{i : c \in \mathcal{C}_i}|$.
Capabilities are ranked by frequency and partitioned into $T$ release buckets, which induces a release stage $r(c)$ for each capability. Frequently used capabilities are introduced earlier, forming the \emph{core}, while less frequent capabilities are released progressively toward the \emph{long tail}. This ordering reflects a common deployment pattern in which broadly useful capabilities are exposed before more specialized ones.

\textbf{Step 3: Cumulative harness construction.} \quad
The harness available at stage $t$ is defined as
\begin{equation}
\mathcal{H}_t \coloneqq \{c \in \mathcal{U} : r(c) \le t \}.
\end{equation}
By construction, the harness grows monotonically, i.e., $\mathcal{H}_t \subseteq \mathcal{H}_{t+1}$, and capabilities are never removed or modified after they are introduced.

\textbf{Step 4: Task assignment.} \quad
Each task is assigned to the earliest stage at which all of its required capabilities are available:
\begin{equation}
t_i \coloneqq \min\{t : \mathcal{C}_i \subseteq \mathcal{H}_t\} = \max_{c \in \mathcal{C}_i} r(c).
\end{equation}
This directly guarantees the two properties stated in \cref{sec:benchmark_overview}: \emph{feasibility} and \emph{new-capability pressure}. The overall pipeline is deterministic, requires neither LLM-generated tasks nor manual verification, and applies to any benchmark with verifier-checkable tasks and capability annotations.

\subsection{Axis-Specific Construction}
\label{sec:axes}

\textbf{Tools.}\quad
The tool axis is the most direct: each task’s capability annotation $\mathcal{C}_i^{\mathrm{tool}}$ is given by the oracle tool set from the source benchmark. Applying the construction pipeline yields a nested sequence of tool catalogs,
$\mathcal{H}_1^{\mathrm{tool}} \subseteq \cdots \subseteq \mathcal{H}_T^{\mathrm{tool}}$.
At stage $t$, the agent receives the full cumulative tool catalog rather than an oracle-pruned task-specific subset, so later stages introduce both newly useful tools and a growing set of potential distractors.

\textbf{Skills.}\quad
Unlike tools, procedural skills are not typically annotated in the source benchmarks. We therefore construct $\mathcal{C}_i^{\mathrm{skill}}$ deterministically in two steps:
\begin{enumerate}[nosep,leftmargin=24pt]
\item \textit{Mining reference skills.} Source system prompts typically contain both general instructions (e.g., role definitions and output-format constraints) and \textbf{reusable step-by-step procedures} for specific operations. We separate these components: general instructions remain in the prompt visible to evaluated systems, while procedural content is extracted into a set of hidden reference skills. This extraction is rule-based rather than LLM-generated.
\item \textit{Associating skills with tasks.} Task verifiers operate on final states (e.g., database entries or file contents). We extract keywords from these verifier-checked states and match them against the hidden reference skills. A skill is associated with a task when entities or identifiers appearing in the verifier-checked state also appear in the skill's procedural content. The resulting matched set defines $\mathcal{C}_i^{\mathrm{skill}}$, which we refer to as the task's \textit{essential-skill annotations}. This keyword-based association is a heuristic proxy rather than an exact correspondence: it identifies skills related to the verifier-checked outcome, but does not guarantee that the matched skills are necessary, sufficient, or unique for solving the task.
\end{enumerate}

Applying the frequency-ranked release construction (\cref{sec:construction}) yields 
$\mathcal{H}_1^{\mathrm{skill}} \subseteq \cdots \subseteq \mathcal{H}_T^{\mathrm{skill}}$.
At each stage, the mined skills form the retrievable skill pool, but are not directly injected into the agent prompt. Instead, the original procedural content is removed from the prompt, requiring the system to identify and retrieve relevant skills from the expanding pool rather than rely on prompt-embedded procedures. Full details of the skill-mining and task-matching procedure are provided in \cref{sec:skill_construction_details}.

\textbf{Agents.}\quad
The agent axis transforms the tool structure into an expanding pool of specialist agents. We define an ownership map $\eta : \mathcal{U}^{\mathrm{tool}} \to \mathcal{E}$ that assigns each tool to the entity it operates on (e.g., the database table storing user or case records, 
or stateful object
such
as a file or email record, as specified by its interface). Tools with the same owner are grouped into a bundle, and each bundle defines an entity-scoped specialist agent $g_e$, , such as a database specialist
$g_{\mathrm{database}}$, file-system specialist $g_{\mathrm{filesystem}}$,
or email specialist $g_{\mathrm{email}}$, 
equipped with the corresponding tools and relevant procedural context identified during skill mining. Additional construction details are provided in \cref{sec:agent_construction_detail}.

A task's agent annotation is defined as $
\mathcal{C}_i^{\mathrm{agent}} \coloneqq \{g_{\eta(u)} \mid u \in \mathcal{C}_i^{\mathrm{tool}}\}.
$ When the tools required by a task span multiple entities (e.g., finding a user in the database and then drafting and sending that user
an email),
the task therefore structurally requires delegation across multiple specialist agents. Applying the frequency-ranked release construction (\cref{sec:construction}) to these task-level annotations yields $
\mathcal{H}_1^{\mathrm{agent}} \subseteq \cdots \subseteq \mathcal{H}_T^{\mathrm{agent}}.$
During evaluation, a lead agent with no direct tool access receives the full cumulative agent pool $\mathcal{H}_t^{\mathrm{agent}}$ and must identify, delegate to, and coordinate the appropriate specialists to solve each task.

\subsection{Evaluation Protocol}
\label{sec:evaluation_protocol}

Given a harness stream $\mathcal{H}_{1:T}$ and its associated task sets, we evaluate systems under two complementary modes that isolate distinct aspects of performance under harness evolution. We report task-level performance and operating cost, together with trajectory-level transfer metrics that distinguish adaptation to newly introduced tasks from retention of previously introduced tasks. Specifically, \textbf{Pass} is the fraction of tasks whose verifier
reports successful completion, while \textbf{Score} is the average verifier
score returned by the source benchmark, capturing partial credit when available. We use relative Forward Transfer (\textbf{FWT}) to measure adaptation to the current-stage cohort and relative Backward Transfer (\textbf{BWT}) to measure changes in performance on earlier cohorts as the harness expands~\citep{wang2024comprehensive}. In the following, we define the two evaluation modes. Full metric definitions and a visual summary of the evaluation modes are provided in Appendix~\ref{sec:metrics}.

\textbf{Deployment evaluation:}
the system carries no persistent state across stages. At each stage $t$, we instantiate a fresh system with harness $\mathcal{H}_t$ and evaluate it on the cumulative evaluation set $
\mathcal{D}_{\le t,\mathrm{eval}} \coloneqq \bigcup_{\tau=1}^{t}\mathcal{D}_{\tau,\mathrm{eval}}.
$
For task $x_i$, we denote the system output by $\mathcal{A}(x_i;\mathcal{H}_t)$. This setting isolates the \emph{direct effect of harness expansion}: whether an agent can exploit newly available capabilities while remaining robust to the growing pool of alternatives, without any benefit or interference from accumulated experience. Any degradation observed in this mode is therefore \textbf{harness-induced}: the model parameters and prompting remain fixed, and only the exposed capability set changes.

\textbf{Self-evolving adaptation evaluation:}
the system maintains a persistent adaptive state $z_t$ that evolves across stages. While $\mathcal{H}_t$ denotes the externally supplied harness, $z_t$ captures the agent's accumulated knowledge for operating within that harness, such as episodic memories of prior tool-use patterns, self-generated procedural skills, or learned routing policies for delegating to specialist agents. We denote the system output at stage $t$ by $\mathcal{A}(x_i; \mathcal{H}_t, z_t)$. Before evaluation at each stage, the system may use the cumulative adaptation set $ \mathcal{D}_{\le t,\mathrm{adapt}} \coloneqq \bigcup_{\tau=1}^{t}\mathcal{D}_{\tau,\mathrm{adapt}}$ to update $z_t$. The specific adaptation strategies considered are described in \cref{sec:experiment}. This setting measures whether accumulated experience supports adaptation to newly introduced capabilities while preserving previously accessible competence, or instead becomes stale, misleading, or harmful as the harness evolves.

\section{Main Results and Analysis}
\label{sec:experiment}
We organize the results by evolution axis—\textit{tools}, \textit{skills}, and \textit{agents}—and progress from deployment and self-evolving adaptation to transfer and mechanism analyses. Under \textbf{deployment evaluation}, we consider representative SoTA \textbf{single-agent systems (SAS)}—ReAct (GPT-5), Codex, and Claude Code (Sonnet-4.6)\footnote{Claude Code uses a different underlying model and harness from the controlled GPT-5-based systems. We therefore omit it from the main comparison tables and report its full results in Appendix~\ref{sec:claude_code_result}.}—as well as \textbf{multi-agent systems (MAS)} including AutoGen~\citep{wu2024autogen} and DeLM~\citep{mao2026decentralized}. For \textbf{self-evolving adaptation}, we instantiate the persistent state $z_t$ (see \cref{sec:evaluation_protocol}) using three families of methods: \textbf{memory-based} approaches (Raw Memory\footnote{Raw Memory is a minimal baseline that directly retains past interaction trajectories without consolidation or learned transformation.}, ReasoningBank~\citep{ouyang2026reasoningbank}, MemToolAgent~\citep{er2026memtoolagent}, G-Memory~\citep{zhang2025g-memory}, and LEGOMem~\citep{han2025legomemmodularproceduralmemory}), which store and retrieve episodic experience; \textbf{prompt-based} adaptation (GEPA~\citep{agrawal2026gepa}), which updates persistent textual instructions from task feedback; and \textbf{code-based} adaptation (Meta-Harness~\citep{lee2026metaharnessendtoendoptimizationmodel}), which optimizes the executable harness itself. Within each comparison, we hold the underlying model and execution setting fixed so that observed differences can be attributed to harness evolution and the adaptation strategy. All evaluations also include a \textbf{task-specific reference} in which each task is exposed only to its annotated required capabilities (\cref{sec:construction}), rather than the full cumulative harness. This provides a controlled comparison between task-specific capability exposure and the broader cumulative harness. We refer to the resulting evolving-harness instances as
\ourbenchmark-EOG and \ourbenchmark-ALE;
in the results section, we shorten these to EOG and ALE when unambiguous.

\subsection{Evolving Tools}
\label{sec:evolving_tools_results}

\begin{table}[t]
\caption{
\small Main results on \textbf{evolving tools}. ``Task-specific tools'' is a \textbf{reference condition} (controlled setting) where each task receives only its annotated required tool set. All other rows use the full cumulative catalog. ReAct is used on EOG and Codex on ALE; \faIcon{network-wired} denotes multi-agent systems. All systems use GPT-5 as the underlying model to enable
controlled comparison across harness and adaptation methods.
}
\vspace{-.8\baselineskip}
\centering
{%
\setlength{\dashlinedash}{0.7pt}%
\setlength{\dashlinegap}{1.2pt}%
\setlength{\arrayrulewidth}{0.2pt}%
\resizebox{\textwidth}{!}{%
\begin{tabular}{
  @{\hskip 4pt}l@{\hskip 6pt}
  cccc@{\hskip 7pt}:
  cccc@{\hskip 6pt}:
  c@{\hskip 4pt}
}
\toprule
& \multicolumn{4}{c}{{\large\textsf{\textbf{\ourbenchmark-EOG}}} (454 tasks)}
& \multicolumn{4}{c}{{\large\textsf{\textbf{\ourbenchmark-ALE}}} (63 tasks)}
& \textbf{Overall} \\
& Pass (\%) & Score & \faIcon[regular]{clock} (h) & Tok. (M)
& Pass (\%) & Score & \faIcon[regular]{clock} (h) & Tok. (M)
& \textbf{Pass Rate} \\
\midrule
\rowcolor{oracletrow}
\multicolumn{10}{l}{%
  \textbf{(Task-specific Reference) Frontier Deployment Systems}
} \\
ReAct / Codex & \hscb{26.0}$^{\pm0.6}$ & \hscb{57.0}$^{\pm1.6}$ & 48.3 & 27.2 & \hscb{11.6}$^{\pm1.5}$ & \hscb{34.6}$^{\pm3.4}$ & 9.7 & 82.2 & \hsc{24.2} \\
\midrule
\rowcolor{agentrow}
\multicolumn{10}{l}{%
  \textbf{Frontier Deployment Systems}
} \\
ReAct / Codex & \hscb{30.2}$^{\pm0.8}$ & \hscb{60.0}$^{\pm0.9}$ & 28.6 & 75.8 & \hscb{12.2}$^{\pm0.7}$ & \hscb{30.0}$^{\pm1.7}$ & 9.2 & 100.8 & \hsc{28.1} 
\\
\faIcon{network-wired}\,AutoGen
& \hscb{30.8}$^{\pm0.1}$ & \hscb{62.6}$^{\pm0.2}$ & 24.4 & 63.7 & \hscb{11.3}$^{\pm2.3}$ & \hscb{32.2}$^{\pm2.4}$ &
  6.5 & 19.6 & \hsc{28.6} \\

\midrule
\rowcolor{modelrow}
\multicolumn{10}{l}{%
  \textbf{Memory-based Self-evolving Adaptation}
} \\
Raw Memory & \hscb{33.0}$^{\pm0.9}$ & \hscb{66.2}$^{\pm0.6}$ & 25.7 & 124.5 & \hscb{7.9}$^{\pm2.2}$ & \hscb{21.4}$^{\pm3.6}$ & 8.4 & 184.3 & \hsc{29.9} \\
Reasoning Bank & \hscb{36.9}$^{\pm1.3}$ & \hscb{68.7}$^{\pm0.4}$ & 21.1 & 150.8 & \hscb{11.1}$^{\pm1.3}$ & \hscb{28.4}$^{\pm1.5}$ & 10.0 & 106.8 & \hsc{33.8}  \\
MemToolAgent & \hscb{38.6}$^{\pm1.0}$ & \hscb{68.9}$^{\pm0.5}$ & 23.9 & 148.0 & \hscb{10.1}$^{\pm2.0}$ & \hscb{25.8}$^{\pm1.2}$ & 13.1 & 83.8 & \textbf{\hsc{35.1}}  \\

\faIcon{network-wired}\,G-Memory
& \hscb{32.2}$^{\pm1.2}$ & \hscb{64.5}$^{\pm0.6}$ & 30.5 & 66.0 & \hscb{13.9}$^{\pm0.1}$ & \hscb{32.5}$^{\pm1.6}$ &
  7.4 & 20.9 & \hsc{30.1} \\

\faIcon{network-wired}\,LegoMem
& \hscb{22.2}$^{\pm0.7}$ & \hscb{57.3}$^{\pm0.4}$ & 52.8 & 141.7 & \hscb{8.3}$^{\pm1.7}$ & \hscb{23.7}$^{\pm2.7}$ &
  13.8 & 39.5 & \hsc{20.6} \\

\faIcon{network-wired}\,DeLM
& \hscb{20.3}$^{\pm1.8}$ & \hscb{51.4}$^{\pm1.5}$ & 97.7 & 179.1 & \hscb{6.2}$^{\pm1.8}$ & \hscb{23.7}$^{\pm1.5}$ &
  16.7 & 30.9 & \hsc{18.7} \\
\midrule
\rowcolor{termrow}
\multicolumn{10}{l}{%
  \textbf{Prompt-based Self-evolving Adaptation}
} \\
GEPA & \hscb{31.9}$^{\pm0.8}$ & \hscb{65.9}$^{\pm0.3}$ & 31.9 & 100.2 & \hscb{11.1}$^{\pm0.0}$ & \hscb{30.9}$^{\pm1.4}$ & 9.6 & 86.9 & \hsc{29.3} \\
\midrule
\rowcolor{harnessrow}
\multicolumn{10}{l}{%
  \textbf{Code-based Self-evolving Adaptation}
} \\
Meta-Harness & \hscb{35.2}$^{\pm0.7}$ & \hscb{65.8}$^{\pm1.1}$ & 27.3 & 99.0 & \hscb{11.1}$^{\pm1.3}$ & \hscb{30.4}$^{\pm1.3}$ & 11.2 & 107.8 & \hsc{32.3} \\
\bottomrule
\end{tabular}}%
\par}
\vspace{-1.5\baselineskip}
\label{tab:tool_main_results}
\end{table}

\textbf{\faIcon{tools}\,Obs.~\ding{182} (Deployment): Broader tool exposure improves accuracy but raises cost.} As shown in \cref{tab:tool_main_results}, compared with the task-specific reference, exposing frontier agents to the full cumulative tool catalog surprisingly improves pass rate on both EOG and ALE, but substantially
increases token usage. Thus, broader tool access can improve deployment accuracy
while making execution considerably more expensive. This suggests that the larger catalog is not merely a source of distraction: agents can benefit from additional affordances, but only through substantially more search and execution.

\textbf{\faIcon{tools}\,Obs.~\ding{183} (Self-evolving adaptation): Additional gains are possible but inconsistent.}
\cref{tab:tool_main_results} further shows that, on EOG, several methods substantially outperform the cumulative deployment baseline: MemToolAgent reaches $38.6\%$ pass rate, followed by ReasoningBank ($36.9\%$) and Meta-Harness ($35.2\%$), compared with $30.2\%$ without adaptation. On ALE, however, most methods remain close to or below the deployment baseline. This contrast shows that accumulating persistent experience does not reliably translate into better performance under harness evolution; its benefit varies substantially across environments.

\begin{wrapfigure}{r}{0.5\columnwidth}
\centering
\includegraphics[width=\linewidth]{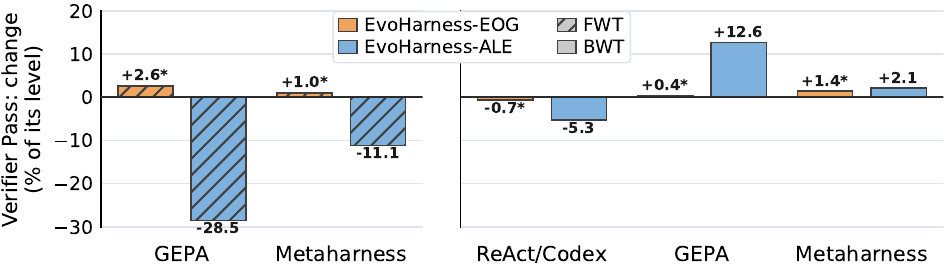}
\vspace{-1.6\baselineskip}
\captionof{figure}{\small Average FWT and BWT on evolving tools. $*$ indicates values disagree among domains. 
}
\label{fig:fwt_bwt_tool}
\vspace{-1\baselineskip}
\end{wrapfigure}

\textbf{\faIcon{tools}\,Obs.~\ding{184} (Transfer): Better retention can come with worse new-task adaptation.} \cref{fig:fwt_bwt_tool} shows substantial \emph{harness-induced forgetting} in the absence of self-evolving adaptation, with deployment BWT negative on both EOG and ALE (see ReAct/Codex). Self-evolving adaptation can improve retention— both GEPA and Meta-Harness achieve positive BWT—but this comes with a sharp FWT degradation in ALE, which falls to $-28.5\%$ and $-11.1\%$, respectively. Since the inner adaptation is explicitly designed to improve performance on newly introduced tasks, this negative FWT is especially notable. It suggests that persistent adaptation may over-specialize to the observed adaptation experience or otherwise interfere with generalization to held-out tasks under the broader tool harness.{Results for MAS follow similar pattern}, as provided in \cref{app:mas_tool_results}.

\begin{wraptable}{r}{0.5\columnwidth}
\caption{\small Effect of adaption context for MemToolAgent on \ourbenchmark-EOG. All conditions follow the same task stream and are evaluated under the cumulative tool harness; ``w/o broad exposure'' restricts each adaptation task
to its task-specific required tools; ``w/o evolution'' exposes the final-stage (total) tool pool at every adaptation stage.
}
\vspace{-1.0em}
\centering
\small
\setlength{\tabcolsep}{4pt}
\resizebox{\linewidth}{!}{%
\begin{tabular}{l l c c c}
\toprule
\textbf{Adapt.} & \textbf{Test} &
\textbf{Pass} & \textbf{Score} & \textbf{Time} \\
\textbf{Harness} & \textbf{Harness} &
\textbf{(\%)} & & \textbf{(h)} \\
\midrule
Cumulative    & Cumulative & 38.6 & 68.9 & 23.9 \\
$\;$ w/o broad tool set & Cumulative  & \textbf{39.1} & \textbf{69.0} & \textbf{17.9} \\
$\;$ w/o evolution   & Cumulative & 37.9 & 68.8 & 28.6 \\
\bottomrule
\end{tabular}
}
\label{tab:tool_adaptation_harness}
\vspace{-1.0em}
\end{wraptable}

\textbf{\faIcon{tools}\,Obs.~\ding{185} (Adaptation under tool-set variations): The harness exposed during adaptation shapes learning.} 
We ask whether the challenge arises only from operating under the expanded
tool catalog at test time, or also from learning persistent state while that
broader harness is exposed during adaptation. 
As shown in \cref{tab:tool_adaptation_harness}, restricting MemToolAgent {(the best-performing method in \cref{tab:tool_main_results})} to task-specific tools during adaptation leaves final performance nearly unchanged, while substantially reducing adaptation time from $23.9$ to $17.9$, despite evaluation under the same cumulative tool harness. In contrast, removing harness evolution while exposing the
final-stage tool pool from the outset provides no benefit: pass rate slightly
decreases to $37.9\%$, while adaptation time increases to $28.6$ hours. However, this pattern is method-dependent: when Raw Memory is likewise adapted using only task-specific tools and
then evaluated under the cumulative tool harness, its final performance improves
more clearly (\cref{tab:tool_adaptation_harness_raw}). These suggest that narrower, task-relevant exposure can make adaptation more efficient, and that the main difficulty lies in adapting under a broad tool space rather than in the stage-wise evolution of that space alone. 

\textbf{\faIcon{tools}\,Obs.~\ding{186} (Impact of Memory): The value of past experience depends on how it shapes search over the expanding tool space.} 
Memory-based methods provide several of the strongest gains in
\cref{tab:tool_main_results}, motivating us to examine how retained experience
affects adaptation under the broader tool harness. For SAS, structured memories
outperform raw replay (MemToolAgent: $38.6\%$; ReasoningBank: $36.9\%$; raw
memory: $33.0\%$), while adding explicit tool schemas to ReasoningBank reduces
performance to $30.3\%$; success-only and failure-only trajectories perform
similarly ($34.4\%$). For MAS, the effect is architecture-dependent: concrete
trajectories can narrow tool search in AutoGen, whereas abstract or broadly
shared memories can increase exploration, redundancy, and cost, and richer
memory can even hurt DeLM. Together, these results suggest that effective memory under tool evolution should extract actionable experience that helps narrow search over the expanding tool space, while avoiding representations that introduce unnecessary exploration or coordination overhead. Full ablations are provided in
Appendix~\ref{sec:evovle_tool_details}.

\subsection{Evolving Skills}
\label{sec:evolving_skills_results}

\begin{table}[t]
\caption{\small Main results on \textbf{evolving skills}.  
We include only SAS, as existing work benchmarking MAS in evolving skills settings remains limited. All systems use the GPT-5 model.
}
\vspace{-.8\baselineskip}
\centering
{%
\setlength{\dashlinedash}{0.7pt}%
\setlength{\dashlinegap}{1.2pt}%
\setlength{\arrayrulewidth}{0.2pt}%
\resizebox{\textwidth}{!}{%
\begin{tabular}{
  @{\hskip 4pt}l@{\hskip 6pt}
  cccc@{\hskip 7pt}:
  cccc@{\hskip 6pt}:
  c@{\hskip 4pt}
}
\toprule
& \multicolumn{4}{c}{{\large\textsf{\textbf{\ourbenchmark-EOG}}} (148 tasks)}
& \multicolumn{4}{c}{{\large\textsf{\textbf{\ourbenchmark-ALE}}} (68 tasks)}
& \textbf{Overall} \\
& Pass (\%) & Score & \faIcon[regular]{clock} (h) & Tok. (M)
& Pass (\%) & Score & \faIcon[regular]{clock} (h) & Tok. (M)
& \textbf{Pass Rate} \\
\midrule
\rowcolor{oracletrow}
\multicolumn{10}{l}{%
  \textbf{(Task-specific Reference) Frontier Deployment System}
} \\
Codex & \hscb{18.9}$^{\pm1.1}$ & \hscb{57.8}$^{\pm0.2}$ & 3.8 & 29.9 & \hscb{7.4}$^{\pm2.1}$ & \hscb{27.9}$^{\pm1.3}$ & 11.4 & 80.1 & \hsc{15.3} \\

\midrule
\rowcolor{agentrow}
\multicolumn{10}{l}{%
  \textbf{Frontier Deployment Systems}
} \\
Codex & \hscb{18.9}$^{\pm0.6}$ & \hscb{59.1}$^{\pm1.1}$ & 3.7 & 31.0 & \hscb{8.3}$^{\pm0.7}$ & \hscb{25.5}$^{\pm0.5}$ & 10.8 & 82.7 & \hsc{15.6} \\
\midrule
\rowcolor{harnessrow}
\multicolumn{10}{l}{%
  \textbf{Memory-based Self-evolving Adaptation}
} \\
Codex Memory & \hscb{17.8}$^{\pm2.2}$ & \hscb{58.0}$^{\pm1.0}$ & 4.6 & 41.5 & \hscb{9.3}$^{\pm1.8}$ & \hscb{25.6}$^{\pm3.5}$ & 11.6 & 87.8 & \hsc{15.1} \\
Raw Memory & \hscb{21.6}$^{\pm1.0}$ & \hscb{61.5}$^{\pm0.3}$ & 2.8 & 93.0 & \hscb{7.4}$^{\pm0.0}$ & \hscb{20.2}$^{\pm1.6}$ & 12.2 & 156.4 & \hsc{17.1}\\

Reasoning Bank & \hscb{20.9}$^{\pm2.0}$ & \hscb{60.2}$^{\pm1.2}$ & 3.2 & 76.1 & \hscb{8.3}$^{\pm1.8}$ & \hscb{23.4}$^{\pm4.1}$ & 12.2 & 82.3 & \hsc{16.9}\\
MemToolAgent & \hscb{19.6}$^{\pm0.6}$ & \hscb{59.2}$^{\pm0.5}$ & 2.9 & 97.0 & \hscb{6.9}$^{\pm2.5}$ & \hscb{22.6}$^{\pm3.7}$ & 11.6 & 79.6 & \hsc{15.6}\\
\midrule
\rowcolor{harnessrow}
\multicolumn{10}{l}{%
  \textbf{Prompt-based Self-evolving Adaptation}
} \\
GEPA & \hscb{24.1}$^{\pm1.4}$ & \hscb{61.0}$^{\pm0.8}$ & 4.1 & 44.2 & \hscb{7.4}$^{\pm2.1}$ & \hscb{27.8}$^{\pm0.6}$ & 11.7 & 97.8 & \textbf{\hsc{18.8}} \\
\midrule
\rowcolor{termrow}
\multicolumn{10}{l}{%
  \textbf{Code-based Self-evolving Adaptation}
} \\
Meta-Harness & \hscb{19.4}$^{\pm1.9}$ & \hscb{59.5}$^{\pm1.0}$ & 4.0 & 36.5 & \hscb{7.8}$^{\pm1.8}$ & \hscb{26.1}$^{\pm2.4}$ & 11.3 & 89.9 & \hsc{15.7} \\
\bottomrule
\end{tabular}}%
\par}
\vspace{-1.5\baselineskip}
\label{tab:skill_main_results}
\end{table}

\textbf{\faIcon{book}\,Obs.~\ding{182} (Deployment): Broader skill exposure has little aggregate effect.} In contrast to evolving tool libraries, moving from task-specific to cumulative skill exposure has almost no effect on deployment performance (\cref{tab:skill_main_results}). On EOG, Codex remains at a $18.9\%$ pass rate, with only a modest increase in token usage. This suggests that expanding the skill library introduces little direct interference, likely because skills must be explicitly retrieved and invoked, rather than being continuously exposed to the model as executable actions.


\textbf{\faIcon{book}\,Obs.~\ding{183} (Self-evolving adaptation): Gains are modest and method-dependent.} Given the limited effect of skill expansion itself, the more important question is whether adaptation can learn to use the available skills more effectively. As shown in \cref{tab:skill_main_results}, GEPA delivers the clearest improvement on EOG, increasing pass rate from $18.9\%$ to $24.1\%$, while the other methods yield smaller or less consistent gains across environments. In contrast to tools, where several memory-based methods benefit from accumulated experience, effective skill adaptation appears to depend more strongly on learning \emph{when} and \emph{how} to retrieve and apply the provided skills.


\begin{wrapfigure}{r}{0.5\columnwidth}
\centering
\vspace{-1.5\baselineskip}
\includegraphics[width=\linewidth]{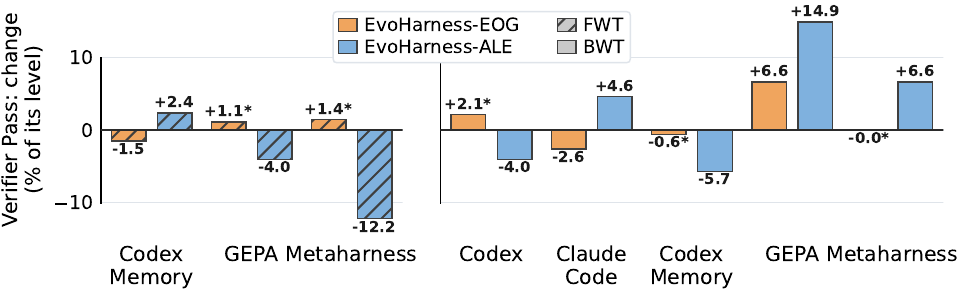}
\vspace{-1.6\baselineskip}
\captionof{figure}{\small Average FWT and BWT on evolving skills. $*$ indicates values disagree among domains.
}
\label{fig:fwt_bwt_skill}
\vspace{-1\baselineskip}
\end{wrapfigure}

\textbf{\faIcon{book}\,Obs.~\ding{184} (Transfer): Mild forgetting persists, while adaptation struggles to exploit newly introduced skills.}
As shown in \cref{fig:fwt_bwt_skill}, deployment exhibits mild forgetting as the skill pool expands {(e.g., Codex on ALE and Claude Code on EOG)}, substantially less than under evolving tools. Self-evolving adaptation can further improve retention—for example, GEPA achieves $+14.9\%$ BWT on ALE—but its FWT remains negative ($-4.0\%$). Thus, preserving competence on earlier tasks does not necessarily translate into effective adaptation to newly introduced skill tasks. Together with the limited effect of skill expansion itself, this suggests a different bottleneck from evolving tools: the challenge lies less in operating with a larger skill library and more in learning persistent updates that reliably retrieve and apply the relevant skills on new tasks.


\begin{wraptable}{r}{0.5\columnwidth}
\caption{\small Effect of adaptation context for GEPA on EOG. We use the same controls as in \cref{tab:tool_adaptation_harness}.
}
\vspace{-1.0em}
\centering
\small
\setlength{\tabcolsep}{4pt}
\resizebox{\linewidth}{!}{%
\begin{tabular}{l l c c c}
\toprule
\textbf{Adapt.} & \textbf{Test} &
\textbf{Pass} & \textbf{Score} & \textbf{Time} \\
\textbf{Harness} & \textbf{Harness} &
\textbf{(\%)} & & \textbf{(h)} \\
\midrule
Cumulative evolving & Cumulative & \textbf{24.1} & 61.0 & 4.1 \\
$\;$ w/o broad skill set
        & Cumulative & 23.0 & \textbf{63.1} & \textbf{2.9} \\
$\;$ w/o evolution     & Cumulative & 23.0 & 59.8 & 7.5 \\
\bottomrule
\end{tabular}
}
\label{tab:skill_adaptation_harness}
\vspace{-1.0em}
\end{wraptable}

\textbf{\faIcon{book}\,Obs.~\ding{185} (Adaptation under skill-set variations): Broad skill exposure can hinder self-evolving adaptation.} 
To probe this difficulty, we vary the harness exposed during adaptation while keeping evaluation under the same cumulative skill pool. \cref{tab:skill_adaptation_harness} reports this analysis for GEPA, the
best-performing method on EOG in \cref{tab:skill_main_results}.
Restricting adaptation to task-specific skills slightly lowers pass rate ($24.1\%$ to $23.0\%$), but improves score ($61.0$ to $63.1$) and substantially reduces adaptation time ($4.1$h to $2.9$h); exposing the final skill pool from the start is still less effective and more costly. This suggests that although a broad skill library is relatively benign at deployment, it can make the inner adaptation problem harder by diluting adaptation experience across potentially irrelevant skills. 

\begin{wrapfigure}{r}{0.5\columnwidth}
\centering
\includegraphics[width=\linewidth]{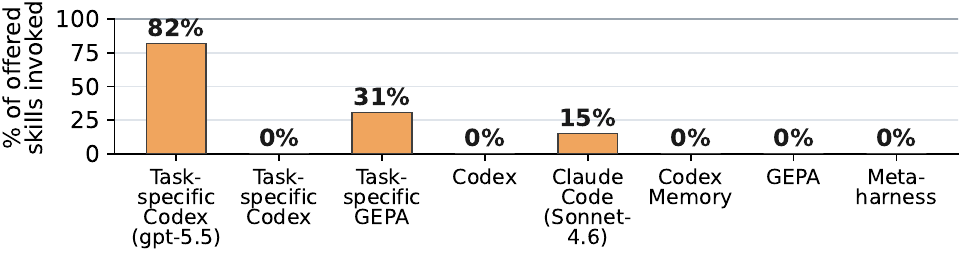}
\vspace{-1.6\baselineskip}
\captionof{figure}{\small Share of offered skills invoked at the final stage across models and adaptation settings. GPT-5 is used unless otherwise specified.
}
\label{fig:skill_invocation_share}
\vspace{-1\baselineskip}
\end{wrapfigure}

\textbf{\faIcon{book}\,Obs.~\ding{186} (Skill engagement): Skill use is sparse, model-dependent, and adaptable.}
Given that reliably selecting and using appropriate skills remains challenging, we further examine whether agents actually engage the skills available to them. As shown in \cref{fig:skill_invocation_share}, default GPT-5-based systems invoke almost none of the offered skills at the final stage, whereas Codex (GPT-5.5) and Claude Code invoke $82\%$ and $15\%$ of the available skills, respectively, in the task-specific setting. Skill engagement can itself be shaped through adaptation: under task-specific training, GEPA increases GPT-5's invocation rate to $31\%$, even when the adapted agent is evaluated against the full skill library. These results indicate that skill utilization depends strongly on both the underlying model and the learned adaptation policy, making effective skill engagement a key bottleneck as the library grows.


\textbf{\faIcon{book}\,Obs.~\ding{187} (Skill learning):
Effective skill acquisition requires grounded refinement, capability coverage, and control over sequential accumulation.}
Beyond adapting through memory, prompts, or code within an externally provided skill harness, many self-evolving systems construct reusable skills from experience. We therefore evaluate skill-learning methods (including Zero-shot and Raw Trajectory, as well as One-shot, Self Feedback, Batch Self
Feedback, Batch Teacher Feedback, and Skill Creator~\citep{zhong2026skilllearnbenchbenchmarkingcontinuallearning})
as a complementary form of self-evolving adaptation under the same evolving-harness protocol. We give details about the methods in Appendix~\ref{app:skill_learn_details}. The results reveal three distinct requirements for effective skill growth. First, \emph{how} skills are learned matters: methods grounded in execution experience and iterative feedback perform best, whereas naive skill generation can underperform using no learned skills at all (\cref{tab:skill_learning_ablation}). Second, improved task performance does not necessarily imply adequate capability coverage, as the generated skills overlap only partially with the annotated skill and subskill space (\cref{fig:skill_iou}). Third, sequential accumulation itself introduces difficulty: even Batch Teacher Feedback, the strongest method in \cref{tab:skill_learning_ablation}, performs better when the same experience is learned jointly than when it is accumulated stage by stage ($23.6\%$ vs.\ $22.1\%$). Together, these results suggest that successful skill learning requires not only generating useful abstractions, but also grounding them in experience, covering the capabilities demanded by the evolving harness, and mitigating interference introduced by sequential updates.


\subsection{Evolving Agents}
\label{sec:evolving_agents_results}

\begin{table}[t]
\caption{\small Main results on \textbf{evolving agents}. 
``--'' denotes experiments that could not be completed within a comparable computational budget. All systems use the GPT-5 model.
}
\vspace{-.8\baselineskip}
\centering
{
\setlength{\dashlinedash}{0.7pt}%
\setlength{\dashlinegap}{1.2pt}%
\setlength{\arrayrulewidth}{0.2pt}%
\resizebox{\textwidth}{!}{%
\begin{tabular}{
  @{\hskip 4pt}l@{\hskip 6pt}
  cccc@{\hskip 7pt}:
  cccc@{\hskip 6pt}:
  c@{\hskip 4pt}
}
\toprule
& \multicolumn{4}{c}{{\large\textsf{\textbf{\ourbenchmark-EOG}}} (148 tasks)}
& \multicolumn{4}{c}{{\large\textsf{\textbf{\ourbenchmark-ALE}}} (63 tasks)}
& \textbf{Overall} \\
& Pass (\%) & Score & \faIcon[regular]{clock} (h) & Tok. (M)
& Pass (\%) & Score & \faIcon[regular]{clock} (h) & Tok. (M)
& \textbf{Pass Rate} \\
\midrule
\rowcolor{oracletrow}
\multicolumn{10}{l}{%
  \textbf{(Task-specific Reference) Frontier Deployment System}
} \\
Codex & \hscb{6.5}$^{\pm1.8}$ & \hscb{36.3}$^{\pm1.6}$ & 24.8 & 570 & \hscb{5.3}$^{\pm3.0}$ & \hscb{19.2}$^{\pm2.9}$ & 27.4 & 20.0 & \hsc{6.2} \\
\midrule
\rowcolor{agentrow}
\multicolumn{10}{l}{%
  \textbf{Frontier Deployment System}
} \\
Codex & \hscb{8.8}$^{\pm4.5}$ & \hscb{42.9}$^{\pm3.0}$ & 25.3 & 587 & \hscb{4.2}$^{\pm2.0}$ & \hscb{16.4}$^{\pm1.3}$ & 37.8 & 30.3 & \hsc{7.4} \\
\midrule
\rowcolor{modelrow}
\multicolumn{10}{l}{%
  \textbf{Memory-based Self-evolving Adaptation}
} \\
Codex Memory & \hscb{10.1}$^{\pm0.0}$ & \hscb{42.3}$^{\pm0.6}$ & 20.2 & 197 & \hscb{3.4}$^{\pm0.7}$ & \hscb{15.4}$^{\pm2.2}$ & 33.5 & 37.4 & \hsc{6.8} \\
G-Memory & \hscb{18.2}$^{\pm1.1}$ & \hscb{55.1}$^{\pm0.7}$ & 29.4 & 30.4 & -- & -- & -- & -- & -- \\
LEGOMemory & \hscb{13.3}$^{\pm0.6}$ & \hscb{44.7}$^{\pm1.4}$ & 16.1 & 21.8 & \hscb{5.5}$^{\pm0.9}$ & \hscb{18.9}$^{\pm2.2}$ &
  13.6 & 34.0 & \hsc{11.0} \\
DeLM  & \hscb{14.2}$^{\pm1.5}$ & \hscb{53.7}$^{\pm1.1}$ & 30.2 & 32.1 & -- & -- & -- & -- & -- \\
\midrule
\rowcolor{harnessrow}
\multicolumn{10}{l}{%
  \textbf{Prompt-based Self-evolving Adaptation}
} \\
GEPA & \hscb{14.9}$^{\pm0.6}$ & \hscb{54.1}$^{\pm0.9}$ & 16.5 & 172 & \hscb{6.3}$^{\pm1.3}$ & \hscb{23.6}$^{\pm0.9}$ & 28.2 & 25.9 & \hsc{12.3} \\
\midrule
\rowcolor{agentrow}
\multicolumn{10}{l}{%
  \textbf{Code-based Self-evolving Adaptation}
} \\
Meta-Harness & \hscb{18.5}$^{\pm2.5}$ & \hscb{57.6}$^{\pm1.2}$ & 17.9 & 228 & \hscb{3.2}$^{\pm0.0}$ & \hscb{18.3}$^{\pm3.0}$ & 40.9 & 29.4 & \textbf{\hsc{13.9}} \\
\bottomrule
\end{tabular}}
\par}
\vspace{-1\baselineskip}
\label{tab:agent_main_results}
\end{table}

\textbf{\faIcon{users}\,Obs.~\ding{182} (Deployment): Agent-pool expansion has strongly environment-dependent effects.} Unlike tools and skills, expanding the agent pool does not induce a consistent deployment trend (\cref{tab:agent_main_results}). Performance is already low under the task-specific reference pool, underscoring the intrinsic difficulty of multi-agent coordination. Moving to the cumulative pool then improves Codex on EOG ($6.5\% \to 8.8\%$) but degrades it on ALE ($5.3\% \to 4.2\%$). Thus, broader access to specialist agents is neither uniformly beneficial nor uniformly benign; its impact depends strongly on the environment-specific demands of agent selection, delegation, and orchestration.


\textbf{\faIcon{users}\,Obs.~\ding{183} (Self-evolving adaptation): Agent evolution yields the largest gains, but they are highly environment-dependent.} On EOG, self-evolving adaptation produces the strongest relative improvements across the three harness axes: Meta-Harness increases pass rate from the $8.8\%$ deployment baseline to $18.5\%$, with several other methods also delivering substantial gains (\cref{tab:agent_main_results}). This suggests that persistent adaptation has particularly high leverage in the more complex agent-coordination setting, where experience can improve selection, delegation, and orchestration. The pattern does not carry over to ALE, however, where most methods remain at or below the $4.2\%$ deployment baseline and only GEPA shows a clear improvement. 


\begin{wrapfigure}{r}{0.5\columnwidth}
\centering
\vspace{-1.6\baselineskip}
\includegraphics[width=\linewidth]{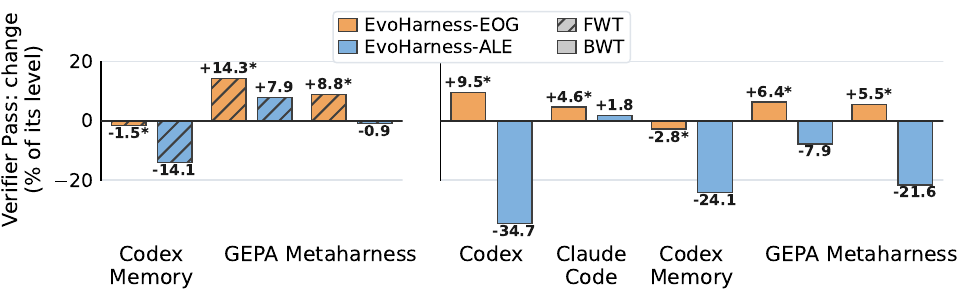}
\vspace{-1.6\baselineskip}
\captionof{figure}{\small Average FWT and BWT on evolving agents. $*$ indicates values disagree among domains.}
\label{fig:fwt_bwt_agent}
\vspace{-1\baselineskip}
\end{wrapfigure}

\textbf{\faIcon{users}\,Obs.~\ding{184} (Transfer): Agent evolution is highly environment-dependent, with severe forgetting on ALE.} As shown in \cref{fig:fwt_bwt_agent}, EOG exhibits relatively favorable transfer: GEPA achieves $+14.3\%^\star$ FWT together with positive BWT, indicating that adaptation can improve performance on newly introduced tasks while preserving competence on earlier cohorts. ALE shows the opposite regime. Most methods exhibit weaker or negative FWT, while deployment Codex reaches $-34.7\%$ BWT, revealing severe harness-induced forgetting from agent-pool expansion alone. This is the strongest forgetting observed across the three harness axes, suggesting that agent evolution is particularly sensitive to environment structure: when specialist capability boundaries are difficult to distinguish, expanding the pool can destabilize previously effective selection and delegation.


\begin{wraptable}{r}{0.5\columnwidth}
\caption{\small Effect of adaptation context for Meta-Harness on \ourbenchmark-EOG. We use the same controls as in \cref{tab:tool_adaptation_harness}.
}
\vspace{-1.0em}
\centering
\small
\setlength{\tabcolsep}{3.5pt}
\resizebox{\linewidth}{!}{%
\begin{tabular}{l l c c c}
\toprule
\textbf{Adapt.} & \textbf{Test} &
\textbf{Pass} & \textbf{Score} & \textbf{Time} \\
\textbf{Harness} & \textbf{Harness} &
\textbf{(\%)} & & \textbf{(h)} \\
\midrule
Cumulative & Cumulative & 18.5 & 57.6 & 17.9 \\
$\;$ w/o broad agent set        & Cumulative & 19.4 & 58.3 & \textbf{15.0} \\
$\;$ w/o  evolution     & Cumulative & \textbf{20.3} & \textbf{60.2} & 17.4 \\
\bottomrule
\end{tabular}
}

\label{tab:agent_adaptation_harness}
\vspace{-1.0em}
\end{wraptable}

\textbf{\faIcon{users}\,Obs.~\ding{185} (Adaptation under agent-set variations): Agent adaptation is sensitive to the harness exposure trajectory.} To isolate the source of this instability, we vary the agent harness presented to Meta-Harness during adaptation while keeping evaluation fixed under the same cumulative pool (\cref{tab:agent_adaptation_harness}). Both controls outperform cumulative evolution, with the largest gain coming from eliminating harness changes altogether: exposing the final agent pool at every stage raises pass rate from $18.5\%$ to $20.3\%$, while task-specific adaptation reaches $19.4\%$. This contrasts with evolving skills, where restricting adaptation to task-relevant capabilities is most beneficial. For agents, a broad pool can still support effective adaptation when it remains stable across stages. These results therefore point to the \emph{changing sequence of available agents}, rather than pool size alone, as a key source of difficulty in learning persistent coordination policies.


\begin{wrapfigure}{r}{0.5\columnwidth}
\centering
\vspace{-0.7\baselineskip}
\includegraphics[width=\linewidth]{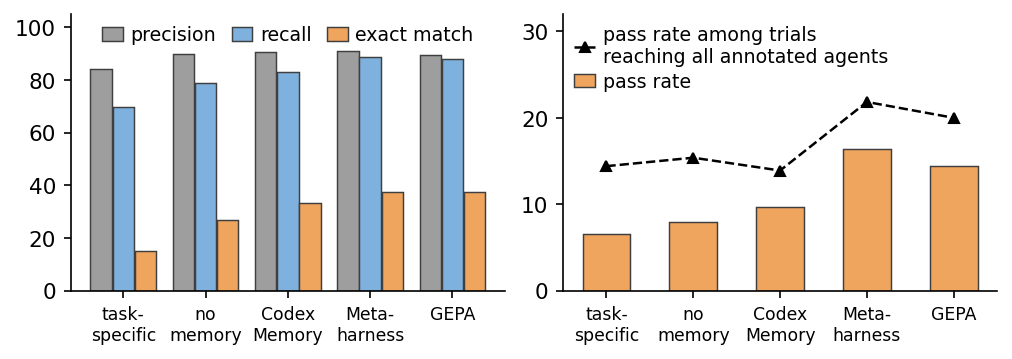}
\vspace{-1.6\baselineskip}
\captionof{figure}{\small Delegation quality.
Left: required-agent precision, recall, and exact match.
Right: overall pass rate and pass rate conditional on reaching all required agents.
}
\label{fig:agent_routing_quality}
\vspace{-0.5\baselineskip}
\end{wrapfigure}

\textbf{\faIcon{users}\,Obs.~\ding{186} (Delegation): Self-evolving adaptation improves delegation completeness more than selection precision.} To understand the source of the large EOG adaptation gains, we examine delegation quality as the agent pool expands (\cref{fig:agent_routing_quality}). Selection precision remains high at roughly $90\%$ across methods, whereas required-agent recall increases from $79\%$ without adaptation to as high as $89\%$, accompanied by higher exact-set coverage. This improved coverage also correlates with stronger end-to-end task success. Thus, self-evolving adaptation appears to help primarily by reaching the full set of required specialists more reliably, rather than by substantially improving judgments of which agents are relevant. A similar pattern emerges for MAS: agent-selection quality is already high across methods, yet better selection accuracy does not consistently translate into higher end-task performance (\cref{fig:mas-agent-selection}). This suggests that, once relevant agents have been identified, the remaining bottleneck lies increasingly in how they are invoked, coordinated, and used together rather than in selection alone.


\begin{wrapfigure}{r}{0.5\columnwidth}
\centering
\includegraphics[width=\linewidth]{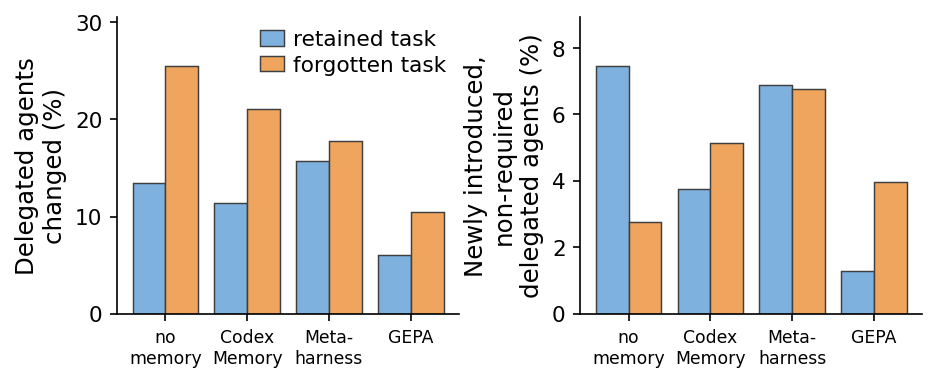}
\vspace{-1.6\baselineskip}
\captionof{figure}{\small Delegation drift on previously solved tasks on \ourbenchmark-EOG. 
Left: change in the delegation set for retained vs.\ forgotten tasks.
Right: share of later delegations to newly introduced, non-required agents.}
\label{fig:agent_routing_drift}
\vspace{-1\baselineskip}
\end{wrapfigure}

\textbf{\faIcon{users}\,Obs.~\ding{187} (Forgetting): Agent-pool expansion can break previously successful delegation patterns.}
To understand how evovling agent can disrupt previously solved tasks,
we re-evaluate previously solved tasks under later agent pools and compare how delegation changes (\cref{fig:agent_routing_drift}). Forgotten tasks show substantially more delegation drift than retained tasks; without adaptation, for example, drift increases from $13\%$ to $25\%$. This is not primarily caused by increased delegation to newly introduced, irrelevant agents. Instead, forgotten tasks more often stop invoking agents that were part of previously successful executions. Thus, harness-induced forgetting appears to arise mainly because agent-pool expansion disrupts effective delegation coverage. For newly introduced agents, the broader challenge is not simply reaching the new specialists, but maintaining coverage of required agents while avoiding unnecessary delegation as the pool grows (\cref{fig:mas-new-specialist}).

\subsection{Summary: Can Agents Keep Pace with an Evolving Harness?}
\label{sec:cross_axis}

Across Sections~\ref{sec:evolving_tools_results}--\ref{sec:evolving_agents_results}, we find that harness evolution affects tools, skills, and agents in distinct ways. Self-evolving adaptation can yield substantial gains, but neither retention of prior competence nor adaptation to newly introduced capabilities is consistently reliable across harness axes and environments. \cref{tab:cross_axis_summary} summarizes these findings. Overall, current agents can benefit from an evolving harness, but they do not yet keep pace with it reliably. The cross-axis results point to three broader lessons about why this remains difficult and what successful adaptation requires.

\begin{table}[h]
\centering
\caption{\small Summary: cross-axis comparison.}
\vspace{-1\baselineskip}
\label{tab:cross_axis_summary}
\small
\renewcommand{\arraystretch}{1.18}
\setlength{\tabcolsep}{3pt}

\resizebox{\columnwidth}{!}{%
\begin{tabular}{lccc}
\toprule
& \textbf{Tools} & \textbf{Skills} & \textbf{Agents} \\
\midrule

\textbf{Deployment expansion}
& Accuracy $\uparrow$, cost $\uparrow$
& Little effect
& Environment-dependent \\

\textbf{Max harness-induced forgetting}
& $-5.3\%$
& $-4.0\%$
& $-34.7\%$ \\

\textbf{Best adaptation gain}
& $+27.8\%$ (MemToolAgent)
& $+27.5\%$ (GEPA)
& $+110.2\%$ (Meta-Harness) \\

\textbf{Best adaptation context}
& Task-specific
& Task-specific
& Final full pool from the beginning \\

\textbf{Observed bottleneck}
& Extract useful tool experience
& Engage the right skills
& Balance delegation coverage \& selectivity \\

\bottomrule
\end{tabular}
}
\end{table}

\textbf{Lesson~\ding{182} (Deployment): Harness evolution itself creates a retention challenge.}
Negative deployment BWT arises even when model parameters and persistent state remain fixed, showing that prior competence can become harder to recover simply because the surrounding harness has expanded. The underlying mechanism differs by axis: tools enlarge the executable action space, skills must be retrieved and engaged, and agents add delegation and coordination demands. Harness-induced forgetting therefore does not stem from a single failure mode; retention must be evaluated under changes to the externally supplied harness itself.

\textbf{Lesson~\ding{183} (Adaptation): Successful adaptation depends on how the harness evolves and what context is available during learning.}
Self-evolving  can yield substantial gains, but the conditions that support those gains differ across harness axes. Tools and skills benefit most from task-relevant exposure, whereas agent adaptation improves more when learning occurs under a stable broader pool. These differences indicate that adaptation quality is shaped not only by the update method itself, but also by the structure and trajectory of the harness presented during learning. Thus, effective inner adaptation must be designed jointly with the outer harness evolution it is expected to track.

\textbf{Lesson~\ding{184} (Transfer): Adaptation gains do not imply retention gains, or vice versa.}
FWT and BWT results expose trade-offs that aggregate performance can obscure. In tools and skills, some adaptation methods improve retention on earlier tasks while reducing performance on newly introduced ones; for agents, adaptation can improve both on EOG yet degrade sharply on ALE. Thus, a system may appear to improve overall while becoming less effective either at exploiting new capabilities or at preserving prior competence. Adaptation and retention should therefore be measured separately throughout the harness trajectory.

\section{Conclusion}

We introduced \ourbenchmark, a benchmark for evaluating agents under
\emph{harness evolution}: controlled growth of the externally supplied tools,
skills, and agents available to a fixed underlying model. By separating an
outer harness evolution from an optional inner self-evolving
adaptation, \ourbenchmark evaluates both deployment under a changing harness and
whether persistent adaptation helps systems keep pace. By evaluating systems
across harness stages rather than only at the final stage,
\ourbenchmark exposes adaptation and retention failures that aggregate
evaluation can obscure.

These results motivate several directions for future work. First, the
\emph{outer harness evolution} and
\emph{inner adaptation} naturally suggests \emph{bi-level adaptation}:
inner adaptation updates optimize persistent state from current adaptation experience,
while an outer objective evaluates whether those updates remain effective across
harness stages, balancing adaptation to newly introduced capabilities with
preservation of earlier competence. Second, the observed retention--adaptation
tension suggests that agents need mechanisms to detect when persistent artifacts
have become stale or harmful and selectively revise, retain, or discard them as
the harness changes. Finally, extending beyond monotonic growth to capability
replacement and retirement would introduce additional migration and
obsolescence challenges that are common in deployed systems.

\bibliography{mas_zero,mas_r1,bgm,evoharness}
\bibliographystyle{salesforce}

\newpage
\appendix
\onecolumn

\counterwithin{table}{section}
\renewcommand{\thetable}{\Alph{section}.\arabic{table}}
\counterwithin{figure}{section}
\renewcommand{\thefigure}{\Alph{section}.\arabic{figure}}

\begin{center}
\textbf{\LARGE Appendix }

\end{center}
\crefalias{section}{appendix} 
\startcontents[appendix]

\printcontents[appendix]{l}{1}{\setcounter{tocdepth}{2}}
\newpage

\section{Contributions}
\label{sec:contributions}

In summary, our key contributions are: 
\begin{itemize}[leftmargin=*,noitemsep,topsep=2pt]

    \item We formalize \textbf{harness evolution} as a distinct source of
    non-stationarity in agentic systems. Unlike conventional continual
    evaluation, where tasks evolve under a fixed harness, we study settings in
    which the externally supplied tools, skills, or agents themselves change
    over time, reshaping what the system observes and what it can do.

    \item We introduce \textbf{\ourbenchmark}, a deterministic framework for
    constructing controlled, cumulative harness streams over \textit{tools},
    \textit{skills}, and \textit{agents} from verifier-based agent benchmarks.
    Capabilities are released from core to long tail, with tasks assigned to the
    earliest feasible stage such that each stage exercises newly introduced
    capabilities, while preserving the original tasks and deterministic evaluation.

    \item We evaluate systems under two complementary modes: \textit{deployment evaluation}, measuring frontier-agent performance as the exposed action space expands, and \textit{self-evolving adaptation evaluation}, measuring whether experience accumulated across harness versions supports adaptation and preservation or instead causes interference. The protocol captures adaptation, retention, transfer, and operating cost along the full harness trajectory.

    \item Through extensive evaluation of SAS and MAS, we reveal
    harness-induced forgetting, substantial variation across capability axes
    and environments, and persistent challenges in learning useful state under
    an expanding harness.

\end{itemize}

\section{Extended Related Work}
\label{sec:relate_detail}

\textbf{Static Harness Benchmarks.}\quad
Existing agent benchmarks evaluate increasingly rich harnesses, with tools, reusable skills, and collaborating agents representing three prominent forms of capability \citep{gu2026modelscalingscalingscaling,galster2026harnessengineeringagenticai}. Tool-use benchmarks evaluate whether agents can select, invoke, and compose executable interfaces \citep{yao2023react,li2023apibankcomprehensivebenchmarktoolaugmented,yao2024taubenchbenchmarktoolagentuserinteraction}; skill-use benchmarks evaluate whether agents can correctly apply provided procedures, scripts, or references \citep{li2026skillsbenchbenchmarkingagentskills,liu2026agenticskillsworkwild,chen2026skillcraftllmagentslearn}; and multi-agent benchmarks evaluate collaboration among predefined roles or specialist agents \citep{zhu2025multiagentbenchevaluatingcollaborationcompetition,liu2026autoresearchclawselfreinforcingautonomousresearch}.

These benchmarks differ in the capabilities they expose, but share a common evaluation assumption: the harness is \emph{fixed} within an evaluation condition. They measure how well an agent performs with a given set of tools, skills, or collaborators, rather than how its performance changes as that capability surface grows over time.

\textbf{Benchmarks for Self-Evolving Agents.}\quad
A growing line of work studies \textit{self-evolving agents} that improve through repeated interactions by updating persistent components such as memory, skills, prompts, workflows, coordination strategies, or harness code \citep{ouyang2026reasoningbank,zhang2025g-memory,hu2024adas,zhang2024aflowautomatingagenticworkflow,ke2025maszero,Ke2026MASOrchestra,lee2026metaharnessendtoendoptimizationmodel,karpathy2026autoresearch}. These methods are commonly evaluated on challenging long-horizon scientific, software, and machine-learning benchmarks, including PaperBench, CORE-Bench, ScienceAgentBench, RE-Bench, MLE-bench, and KernelBench \citep{starace2025paperbenchevaluatingaisability,siegel2026corebenchfosteringcredibilitypublished,chen2025scienceagentbenchrigorousassessmentlanguage,wijk2025rebenchevaluatingfrontierai,chan2025mlebenchevaluatingmachinelearning,ouyang2025kernelbenchllmswriteefficient}. Some self-evolution benchmarks, such as SkillLearnBench and SkillFlow \citep{zhong2026skilllearnbenchbenchmarkingcontinuallearning,zhang2026skillflowbenchmarkinglifelongskilldiscovery}, do involve a growing capability surface as agents generate and refine reusable skills over time. However, this growth is \emph{agent-driven}---a product of the self-evolution process itself---and the evaluation targets skill generation quality rather than how task performance responds to that growth.

Across these benchmarks, the common pattern is that evaluation measures whether self-evolution improves final task performance, but does not track how the agent behaves across the evolution process---whether earlier competence is preserved, or whether accumulated artifacts remain useful as conditions change.

\textbf{Continual Learning Benchmarks.}\quad
Continual-agent benchmarks introduce temporal non-stationarity, but differ in \emph{what} changes over time and \emph{what} that change is used to evaluate. AgentCL constructs controlled task streams to measure transfer and memory-induced interference \citep{shu2026agentclrigorousevaluationcontinual}. ContinualLearningBench uses expert-validated task sequences with shared latent structure \citep{asawa2026continuallearningbenchevaluating}. These benchmarks rigorously evaluate adaptation and preservation across sequential tasks, but the agent's available harness remains fixed. Consequently, they can reveal when accumulated experience becomes harmful as the task distribution shifts, but not when that experience becomes stale because the harness through which the agent acts has changed.

The closest prior setting is CMRBench~\citep{bell2026continual}, which studies continual routing over an externally growing model pool. This is the only existing benchmark where harness growth is environment-driven rather than agent-driven. However, it considers a single axis (models) and evaluates routing decisions only, without end-to-end agentic execution or coordination. \ourbenchmark evaluates end-to-end agent performance under controlled harness growth across tools, skills, and agents.

\section{\ourbenchmark Statistics}
\label{sec:statistics}

\begin{table*}[t]
  \centering
  \caption{Tool-evolution statistics by domain and stage. Percentages are relative to each domain sample; adaptation and test splits are disjoint.}
  \label{tab:eog-tools-evolution}
  \begingroup
  \footnotesize
  \setlength{\tabcolsep}{4.5pt}
  \renewcommand{\arraystretch}{0.96}
  \begin{tabular}{llrrrrrr}
    \toprule
    Domain & Stage & \shortstack{Train\\(adapt)} & Test & \shortstack{Total\\tasks} & \shortstack{\% of\\sample} & \shortstack{+ new\\tools} & \shortstack{cumulative\\tools} \\
    \midrule
    \multirow{4}{*}{\textbf{EOG · Calendar}} & $\mathcal{H}_1$ & 12 & 29 & 41 & 67.2\% & +17 & 17 \\
     & $\mathcal{H}_2$ & 4 & 10 & 14 & 23.0\% & +7 & 24 \\
     & $\mathcal{H}_3$ & 2 & 4 & 6 & 9.8\% & +4 & 28 \\
     & \textbf{Total} & \textbf{18} & \textbf{43} & \textbf{61} & \textbf{100\%} & \textemdash & \textbf{28} \\
    \midrule
    \multirow{5}{*}{\textbf{EOG · CSM}} & $\mathcal{H}_1$ & 5 & 13 & 18 & 17.5\% & +25 & 25 \\
     & $\mathcal{H}_2$ & 7 & 15 & 22 & 21.4\% & +10 & 35 \\
     & $\mathcal{H}_3$ & 9 & 20 & 29 & 28.2\% & +11 & 46 \\
     & $\mathcal{H}_4$ & 10 & 24 & 34 & 33.0\% & +15 & 61 \\
     & \textbf{Total} & \textbf{31} & \textbf{72} & \textbf{103} & \textbf{100\%} & \textemdash & \textbf{61} \\
    \midrule
    \multirow{4}{*}{\textbf{EOG · Drive}} & $\mathcal{H}_1$ & 13 & 29 & 42 & 65.6\% & +22 & 22 \\
     & $\mathcal{H}_2$ & 4 & 10 & 14 & 21.9\% & +6 & 28 \\
     & $\mathcal{H}_3$ & 2 & 6 & 8 & 12.5\% & +8 & 36 \\
     & \textbf{Total} & \textbf{19} & \textbf{45} & \textbf{64} & \textbf{100\%} & \textemdash & \textbf{36} \\
    \midrule
    \multirow{7}{*}{\textbf{EOG · Email}} & $\mathcal{H}_1$ & 1 & 2 & 3 & 4.5\% & +12 & 12 \\
     & $\mathcal{H}_2$ & 3 & 6 & 9 & 13.4\% & +10 & 22 \\
     & $\mathcal{H}_3$ & 3 & 7 & 10 & 14.9\% & +10 & 32 \\
     & $\mathcal{H}_4$ & 5 & 12 & 17 & 25.4\% & +11 & 43 \\
     & $\mathcal{H}_5$ & 6 & 13 & 19 & 28.4\% & +11 & 54 \\
     & $\mathcal{H}_6$ & 3 & 6 & 9 & 13.4\% & +10 & 64 \\
     & \textbf{Total} & \textbf{21} & \textbf{46} & \textbf{67} & \textbf{100\%} & \textemdash & \textbf{64} \\
    \midrule
    \multirow{6}{*}{\textbf{EOG · HR}} & $\mathcal{H}_1$ & 4 & 10 & 14 & 13.7\% & +23 & 23 \\
     & $\mathcal{H}_2$ & 5 & 13 & 18 & 17.6\% & +13 & 36 \\
     & $\mathcal{H}_3$ & 11 & 25 & 36 & 35.3\% & +12 & 48 \\
     & $\mathcal{H}_4$ & 9 & 22 & 31 & 30.4\% & +14 & 62 \\
     & $\mathcal{H}_5$ & 1 & 2 & 3 & 2.9\% & +4 & 66 \\
     & \textbf{Total} & \textbf{30} & \textbf{72} & \textbf{102} & \textbf{100\%} & \textemdash & \textbf{66} \\
    \midrule
    \multirow{5}{*}{\textbf{EOG · Hybrid}} & $\mathcal{H}_1$ & 10 & 24 & 34 & 38.6\% & +82 & 82 \\
     & $\mathcal{H}_2$ & 5 & 11 & 16 & 18.2\% & +25 & 107 \\
     & $\mathcal{H}_3$ & 4 & 9 & 13 & 14.8\% & +25 & 132 \\
     & $\mathcal{H}_4$ & 8 & 17 & 25 & 28.4\% & +31 & 163 \\
     & \textbf{Total} & \textbf{27} & \textbf{61} & \textbf{88} & \textbf{100\%} & \textemdash & \textbf{163} \\
    \midrule
    \multirow{4}{*}{\textbf{EOG · ITSM}} & $\mathcal{H}_1$ & 20 & 48 & 68 & 66.0\% & +39 & 39 \\
     & $\mathcal{H}_2$ & 5 & 12 & 17 & 16.5\% & +10 & 49 \\
     & $\mathcal{H}_3$ & 5 & 13 & 18 & 17.5\% & +15 & 64 \\
     & \textbf{Total} & \textbf{30} & \textbf{73} & \textbf{103} & \textbf{100\%} & \textemdash & \textbf{64} \\
    \midrule
    \multirow{5}{*}{\textbf{EOG · Teams}} & $\mathcal{H}_1$ & 9 & 20 & 29 & 47.5\% & +25 & 25 \\
     & $\mathcal{H}_2$ & 4 & 11 & 15 & 24.6\% & +8 & 33 \\
     & $\mathcal{H}_3$ & 3 & 6 & 9 & 14.8\% & +9 & 42 \\
     & $\mathcal{H}_4$ & 2 & 6 & 8 & 13.1\% & +8 & 50 \\
     & \textbf{Total} & \textbf{18} & \textbf{43} & \textbf{61} & \textbf{100\%} & \textemdash & \textbf{50} \\
    \midrule
    \multirow{6}{*}{\textbf{ALE}} & $\mathcal{H}_1$ & 22 & 52 & 74 & 49.3\% & +45 & 45 \\
     & $\mathcal{H}_2$ & 4 & 8 & 12 & 8.0\% & +27 & 72 \\
     & $\mathcal{H}_3$ & 4 & 8 & 12 & 8.0\% & +27 & 99 \\
     & $\mathcal{H}_4$ & 4 & 8 & 12 & 8.0\% & +28 & 127 \\
     & $\mathcal{H}_5$ & 12 & 28 & 40 & 26.7\% & +51 & 178 \\
     & \textbf{Total} & \textbf{46} & \textbf{104} & \textbf{150} & \textbf{100\%} & \textemdash & \textbf{178} \\
    \bottomrule
  \end{tabular}
  \endgroup
\end{table*}

\begin{table*}[t]
  \centering
  \caption{Skill-evolution statistics by domain and version. Percentages are relative to each domain sample; adaptation and test splits are disjoint.}
  \label{tab:eog-skills-evolution}
  \begingroup
  \footnotesize
  \setlength{\tabcolsep}{4.5pt}
  \renewcommand{\arraystretch}{0.96}
  \begin{tabular}{llrrrrrr}
    \toprule
    Domain & Stage & \shortstack{Train\\(adapt)} & Test & \shortstack{Total\\tasks} & \shortstack{\% of\\sample} & \shortstack{+ new\\skills} & \shortstack{cumulative\\skills} \\
    \midrule
    \multirow{4}{*}{\textbf{EOG · CSM}}
    & $H_1$ & 5 & 12 & 17 & 34.0\% & +2 & 2 \\
    & $H_2$ & 5 & 13 & 18 & 36.0\% & +2 & 4 \\
    & $H_3$ & 4 & 11 & 15 & 30.0\% & +5 & 9 \\
     & \textbf{Total} & \textbf{14} & \textbf{36} & \textbf{50} & \textbf{100\%} & \textemdash & \textbf{9} \\
    \midrule
    \multirow{4}{*}{\textbf{EOG · HR}}
    & $H_1$ & 4 & 11 & 15 & 20.0\% & +2 & 2 \\
    & $H_2$ & 9 & 22 & 31 & 41.3\% & +2 & 4 \\
    & $H_3$ & 9 & 20 & 29 & 38.7\% & +6 & 10 \\
    & \textbf{Total} & \textbf{22} & \textbf{53} & \textbf{75}
    & \textbf{100\%} & \textemdash & \textbf{10} \\
    \midrule
    \multirow{5}{*}{\textbf{EOG · ITSM}}
    & $H_1$ & 5 & 13 & 18 & 21.7\% & +3 & 3 \\
    & $H_2$ & 8 & 20 & 28 & 33.7\% & +2 & 5 \\
    & $H_3$ & 6 & 15 & 21 & 25.3\% & +2 & 7 \\
    & $H_4$ & 5 & 11 & 16 & 19.3\% & +3 & 10 \\
    & \textbf{Total} & \textbf{24} & \textbf{59} & \textbf{83}
    & \textbf{100\%} & \textemdash & \textbf{10} \\
    \midrule
    \multirow{7}{*}{\textbf{ALE}}
    & $H_1$ & 5 & 13 & 18 & 12.4\% & +4 & 4 \\
    & $H_2$ & 4 & 11 & 15 & 10.3\% & +2 & 6 \\
    & $H_3$ & 6 & 15 & 21 & 14.5\% & +1 & 7 \\
    & $H_4$ & 11 & 25 & 36 & 24.8\% & +2 & 9 \\
    & $H_5$ & 8 & 17 & 25 & 17.2\% & +1 & 10 \\
    & $H_6$ & 9 & 21 & 30 & 20.7\% & +3 & 13 \\
     & \textbf{Total} & \textbf{43} & \textbf{102} & \textbf{145} & \textbf{100\%} & \textemdash & \textbf{13} \\ 
    \bottomrule
  \end{tabular}
  \endgroup
\end{table*}

\begin{table*}[t]
  \centering
  \caption{Agent-evolution statistics by domain and version. Percentages are relative to each domain sample; adaptation and test splits are disjoint.}
  \label{tab:eog-agents-evolution}
  \begingroup
  \footnotesize
  \setlength{\tabcolsep}{4.5pt}
  \renewcommand{\arraystretch}{0.96}
  \begin{tabular}{llrrrrrr}
    \toprule
    Domain & Stage & \shortstack{Train\\(adapt)} & Test & \shortstack{Total\\tasks} & \shortstack{\% of\\sample} & \shortstack{+ new\\agents} & \shortstack{cumulative\\agents} \\
    \midrule
    \multirow{4}{*}{\textbf{EOG · CSM}}
     & $H_1$ & 2 & 5 & 7 & 14.0\% & +10 & 10 \\
     & $H_2$ & 2 & 6 & 8 & 16.0\% & +3 & 13 \\
     & $H_3$ & 10 & 25 & 35 & 70.0\% & +5 & 18 \\
     & \textbf{Total} & \textbf{14} & \textbf{36} & \textbf{50} & \textbf{100\%} & \textemdash & \textbf{18} \\
    \midrule
    \multirow{5}{*}{\textbf{EOG · HR}}
     & $H_1$ & 3 & 8 & 11 & 14.7\% & +6 & 6 \\
     & $H_2$ & 2 & 13 & 15 & 20.0\% & +3 & 9 \\
     & $H_3$ & 6 & 17 & 23 & 30.7\% & +3 & 12 \\
     & $H_4$ & 11 & 15 & 26 & 34.7\% & +6 & 18 \\
     & \textbf{Total} & \textbf{22} & \textbf{53} & \textbf{75} & \textbf{100\%} & \textemdash & \textbf{18} \\
    \midrule
    \multirow{5}{*}{\textbf{EOG · ITSM}}
     & $H_1$ & 4 & 7 & 11 & 13.3\% & +4 & 4 \\
     & $H_2$ & 7 & 11 & 18 & 21.7\% & +3 & 7 \\
     & $H_3$ & 3 & 12 & 15 & 18.1\% & +3 & 10 \\
     & $H_4$ & 10 & 29 & 39 & 47.0\% & +5 & 15 \\
     & \textbf{Total} & \textbf{24} & \textbf{59} & \textbf{83} & \textbf{100\%} & \textemdash & \textbf{15} \\
    \midrule
\multirow{7}{*}{\textbf{ALE}}
 & $H_1$ & 12 & 24 & 36 & 24.0\% & +2 & 2 \\
 & $H_2$ & 9 & 18 & 27 & 18.0\% & +3 & 5 \\
 & $H_3$ & 5 & 17 & 22 & 14.7\% & +3 & 8 \\
 & $H_4$ & 10 & 13 & 23 & 15.3\% & +4 & 12 \\
 & $H_5$ & 4 & 17 & 21 & 14.0\% & +4 & 16 \\
 & $H_6$ & 6 & 15 & 21 & 14.0\% & +8 & 24 \\
 & \textbf{Total} & \textbf{46} & \textbf{104} & \textbf{150} & \textbf{100\%} & \textemdash & \textbf{24} \\
    \bottomrule
  \end{tabular}
  \endgroup
\end{table*}

\begin{table}[t]
\centering
\caption{\small
Statistics of \ourbenchmark by harness axis and trajectory.
Each row corresponds to one independent harness stream.
\textit{Sum across streams} sums the final harness sizes of individual trajectories,
whereas \textit{globally unique} deduplicates same-named capabilities across trajectories. \emph{Req./Task} counts oracle tools, verifier-grounded reference skills,
or annotated specialist agents, respectively. \emph{Final-ver.\ distractors} is
the mean fraction of the last version's cumulative capability pool that is
non-oracle for a task.
}
\label{tab:benchmark_stats}
\small
\renewcommand{\arraystretch}{1.12}
\setlength{\tabcolsep}{4pt}
\resizebox{\textwidth}{!}{%
\begin{tabular}{llrrrrr}
\toprule
\textbf{Axis} & \textbf{Trajectory}
& \textbf{Tasks}
& \textbf{Stages}
& \textbf{Final Harness Size}
& \textbf{Req./Task}
& \textbf{Final Distractors} \\
\midrule

\multirow{11}{*}{\textbf{Tools}}
& Calendar & 61  & 3 & 28  & 5.84 & 79.2\% \\
& CSM      & 103 & 4 & 61  & 9.78 & 84.0\% \\
& Drive    & 64  & 3 & 36  & 6.47 & 82.0\% \\
& Email    & 67  & 6 & 64  & 5.69 & 91.1\% \\
& HR       & 102 & 5 & 66  & 8.79 & 86.7\% \\
& Hybrid   & 88  & 4 & 163 & 7.32 & 95.5\% \\
& ITSM     & 103 & 3 & 64  & 6.93 & 89.2\% \\
& Teams    & 61  & 4 & 50  & 8.23 & 83.5\% \\
& ALE      & 150 & 5 & 178 & 3.11 & 98.3\% \\
\cmidrule(lr){2-7}
& \textit{Sum across streams}
& 799 & 37 & 710 & 6.79 & 89.0\% \\
& \textit{Globally unique}
& -- & -- & \textbf{520} & -- & -- \\
\midrule

\multirow{6}{*}{\textbf{Skills}}
& CSM  & 50  & 3 & 9  & 1.70 & 81.1\% \\
& HR   & 75  & 3 & 10 & 2.95 & 70.5\% \\
& ITSM & 83  & 4 & 10 & 3.33 & 66.7\% \\
& ALE  & 145 & 6 & 13 & 4.01 & 69.2\% \\
\cmidrule(lr){2-7}
& \textit{Sum across streams}
& 353 & 16 & 42 & 3.30 & -- \\
& \textit{Globally unique}
& -- & -- & \textbf{42} & -- & -- \\
\midrule

\multirow{6}{*}{\textbf{Agents}}
& CSM  & 50  & 3 & 18 & 7.70 & 57.2\% \\
& HR   & 75  & 4 & 18 & 5.63 & 68.7\% \\
& ITSM & 83  & 4 & 15 & 4.54 & 69.7\% \\
& ALE  & 150 & 6 & 24 & 2.11 & 91.2\% \\
\cmidrule(lr){2-7}
& \textit{Sum across streams}
& 358 & 17 & 75 & 4.19 & 76.8\% \\
& \textit{Globally unique}
& -- & -- & \textbf{62} & -- & -- \\
\bottomrule
\end{tabular}
}
\end{table}

This appendix reports construction statistics and diagnostics for each
\ourbenchmark evolution axis. For every axis, we report both version-level
statistics and construction checks. Version-level statistics characterize how
the external harness grows over time; construction checks verify that each task
is assigned to a valid release timestep and that the evaluated system receives
the intended exposed harness rather than an oracle-pruned capability set.

\textbf{Evolving Tools.}\quad
We report tool-evolution statistics by version, including the number of newly
released tools, cumulative tools, adaptation tasks, evaluation tasks, average
oracle tools per task, and the fraction of exposed tools that are distractors
for a given task. A distractor tool is any tool in the cumulative tool catalog
$\mathcal{H}^{\mathrm{tool}}_t$ that is not in the task's oracle tool set
$\mathcal{C}^{\mathrm{tool}}_i$. We report both the average number and fraction
of distractor tools at each timestep.

In addition to aggregate statistics, we use four construction diagnostics.
First, we verify that each retained task has a nonempty oracle tool annotation
and a verifier-checkable outcome. Second, we verify that the tool catalog
expands according to the release schedule and that
$\mathcal{H}^{\mathrm{tool}}_t \subseteq
\mathcal{H}^{\mathrm{tool}}_{t+1}$ for all $t$. Third, we check that the
release schedule follows the intended core-to-long-tail structure by reporting
tool frequency distributions before and after bucket assignment. Fourth, for
each assigned task, we store the oracle tool configuration and the full exposed
tool catalog at its assigned timestep. These records verify that the oracle
tool set is covered by the exposed harness while ensuring that evaluated
systems receive the full cumulative catalog, not an oracle-pruned tool list.

\textbf{Evolving Skills.}\quad
We report skill-evolution statistics by version, including task coverage, the
number of newly released and cumulative latent skills, adaptation and evaluation
tasks, essential-skill tags per task, and skill reuse frequency at each
timestep. Task coverage is the fraction of seed tasks for which the deterministic
skill matcher assigns at least one verifier-grounded essential-skill tag. Skill
reuse frequency measures how many tasks are tagged with each latent skill and
is used to construct the core-to-long-tail release schedule.

We also report diagnostics for the hidden reference skill construction. First,
we report how much prompt content is retained as model-visible behavioral
contract versus hidden as reusable procedural skill content. Second, we report
the number of mined skill records and their entity, field, value, and outcome
indices. Third, we report verifier-matching coverage: the number of tasks with
at least one matched skill, the distribution of matched skills per task, and the
number of tasks excluded because no deterministic match is found. Fourth, we
verify that the hidden reference skill timeline is used only for task assignment
and evaluation analysis: the evaluated system receives the stripped
agent-facing prompt and seed-annotated tools, but not the hidden reference
skills. These diagnostics ensure that the skill axis evaluates adaptive skill
learning rather than direct retrieval of the reference skill curriculum.

\textbf{Evolving Agents.}\quad
We report agent-evolution statistics by version, including construction
coverage, the number of newly released and cumulative agents, adaptation and
evaluation tasks, agents per task, the fraction of tasks that require multiple
agents, and the distractor-agent fraction at each timestep. A distractor agent
is any specialist agent in the cumulative agent pool
$\mathcal{H}^{\mathrm{agent}}_t$ that is not in the task's annotated agent set
$\mathcal{C}^{\mathrm{agent}}_i$. Multi-agent task fraction measures the share
of tasks whose oracle tools span at least two entity-scoped specialists.

We additionally report construction checks for the specialist-agent pool. First,
we verify that every seed tool is assigned to exactly one owner entity by the
ownership map $\eta:\mathcal{U}^{\mathrm{tool}}\rightarrow\mathcal{E}$.
Second, we verify that every active specialist has a nonempty tool bundle and
that specialist tool bundles are pairwise disjoint. Third, we verify
annotated-tool coverage: for every retained task, the union of the tool bundles
owned by $\mathcal{C}^{\mathrm{agent}}_i$ covers the task's oracle tool set
$\mathcal{C}^{\mathrm{tool}}_i$. Fourth, we report routing complexity through
the number of specialists required per task and the fraction of tasks requiring
coordination across multiple specialists. Finally, we verify the evaluation
exposure: the lead agent receives the full cumulative specialist pool
$\mathcal{H}^{\mathrm{agent}}_t$ and no direct tool access, so success requires
selecting, delegating to, and coordinating the appropriate specialists.

In total, constructed from
two verifier-based agent benchmarks, EnterpriseOps-Gym and Agents' Last Exam,
\ourbenchmark comprises \textbf{802 unique benchmark tasks} instantiated as
\textbf{1{,}510 axis-specific task examples} across 17 independent evolution
trajectories (3--6 cumulative versions each; 70 harness snapshots). These
timelines span \textbf{520 executable tools}, \textbf{42 latent reference
skills}, and \textbf{62 specialist agents}. Skill construction covers 76.6\% of eligible seed tasks, while 85.2\%
of agent-axis tasks require coordination across multiple specialists.
By the final harness stages, each task is evaluated alongside an average
of 89.0\% distractor tools or 76.8\% distractor agents, creating increasing
selection and coordination pressure as capabilities accumulate.

\section{Metrics} 
\label{sec:metrics}

The core evaluation object is a \textbf{lower-triangular performance matrix} $P \in \mathbb{R}^{T \times T}$, where entry $P_{t,\tau}$ ($\tau \le t$) is the accuracy of the system on task cohort $\tau$ when evaluated under harness stage $t$:
\begin{equation}
\label{eq:perf-matrix}
P_{t,\tau} = \frac{1}{|\mathcal{D}_{\tau,\mathrm{eval}}|} \sum_{(x_i,y_i) \in \mathcal{D}_{\tau,\mathrm{eval}}} V\!\left(\mathcal{A}(x_i; \mathcal{H}_t, z_t),\; y_i\right).
\end{equation}
Each row corresponds to a harness stage; each column to a task cohort. The matrix is lower-triangular because cohort $\tau$ only exists once stage $\tau$ is reached.

Different views of this matrix answer different questions:

\textbf{Adaptation (diagonal).} $P_{t,t}$ measures accuracy on newly introduced tasks under their intended harness---can the system exploit newly available capabilities?

\textbf{Retention (off-diagonal row).} $P_{t,\tau}$ for $\tau < t$ measures accuracy on earlier task cohorts under the expanded harness---does the system preserve competence as capabilities accumulate?

\textbf{Backward transfer (column delta).}
\begin{equation}
\mathrm{BWT} = \frac{\sum_{i=1}^{T-1} n_i (P_{T,i} - P_{i,i})}{\sum_{i=1}^{T-1} n_i},
\end{equation}
where $n_i = |\mathcal{D}_{i,\mathrm{eval}}|$. BWT compares each cohort's final-stage performance against its performance at introduction. Negative BWT indicates forgetting: previously accessible competence has degraded as the harness expanded or as persistent state accumulated.

\textbf{Forward transfer (diagonal delta).}
\begin{equation}
\mathrm{FWT} = \frac{\sum_{i=2}^{T} n_i (P_{i,i} - P_{i,i}^{\mathrm{before}})}{\sum_{i=2}^{T} n_i},
\end{equation}
where $P_{i,i}^{\mathrm{before}}$ is cohort $i$'s accuracy under harness $\mathcal{H}_i$ \emph{before} stage-$i$ adaptation (i.e., using $z_{i-1}$). FWT measures how much self-evolution at stage $i$ improves performance on that stage's new tasks. Negative FWT indicates that adaptation actively harms the system's ability to use newly introduced capabilities.

\textbf{Cumulative score.} As an aggregate summary, we report micro-averaged accuracy across all cohorts at the final stage:
\begin{equation}
\label{eq:benchmark-objective}
\mathrm{ACC} = \frac{1}{|\mathcal{D}_{\le T,\mathrm{eval}}|}
\sum_{(x_i,y_i)\in \mathcal{D}_{\le T,\mathrm{eval}}}
V\!\left(\mathcal{A}(x_i;\mathcal{H}_T, z_T),\; y_i\right).
\end{equation}

\textbf{Cost.} We additionally report operating cost (tokens, tool calls, latency) at each stage, capturing efficiency degradation as the harness grows.

Together, the matrix and its derived metrics diagnose \emph{when} harness expansion helps or hurts (which stage), \emph{what} is affected (new vs. old tasks), and \emph{why} (harness-induced distraction vs. stale persistent state).

\begin{figure}[t]
\centering
\includegraphics[width=\columnwidth]{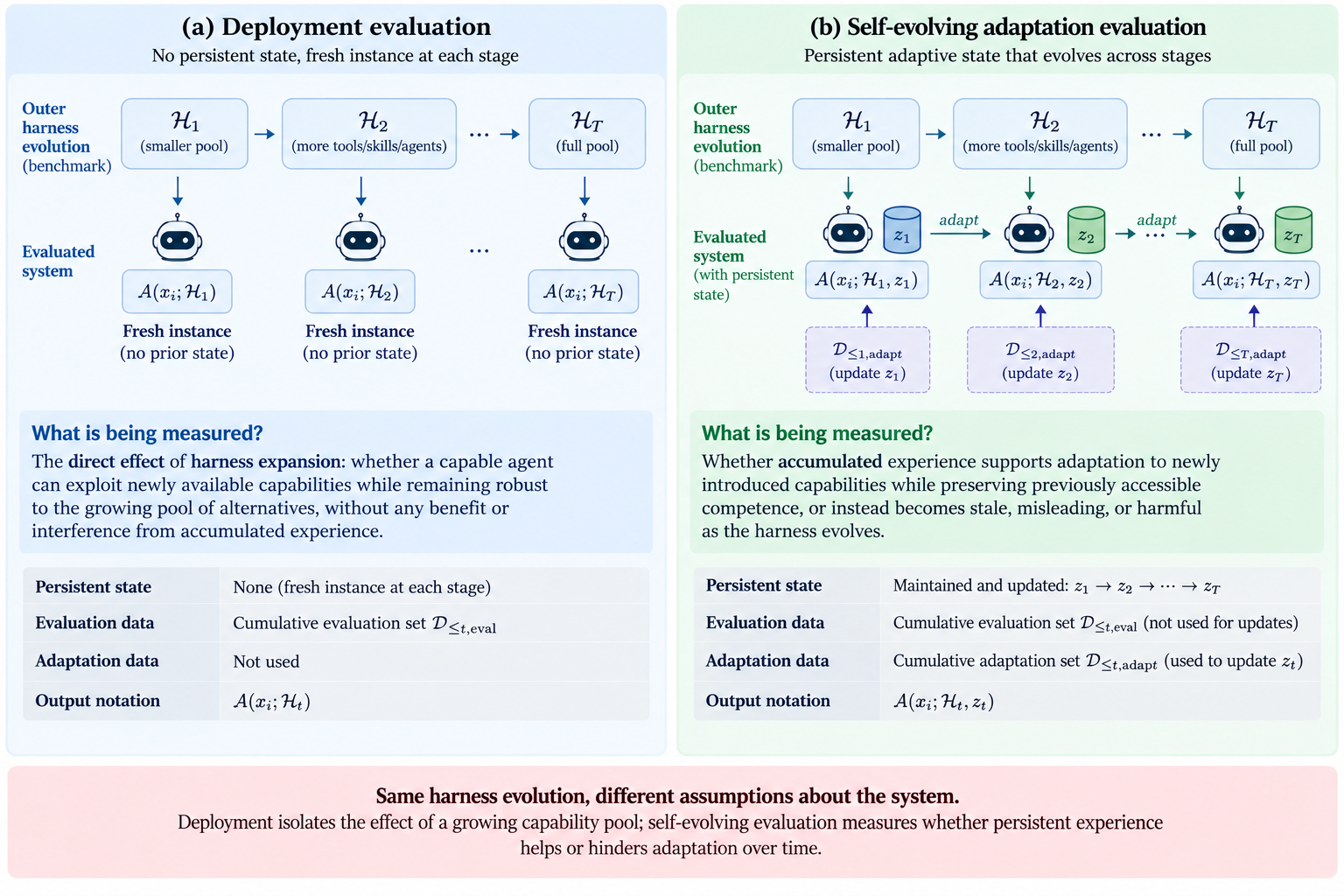}
\vspace{-1\baselineskip}
\captionof{figure}{
Evaluation protocol of \ourbenchmark under the same outer harness evolution
$\mathcal{H}_1 \subseteq \cdots \subseteq \mathcal{H}_T$.
\textbf{(a) Deployment evaluation} uses a fresh system instance at each stage
and evaluates it on the cumulative held-out task set, isolating the direct
effect of harness expansion without accumulated experience.
\textbf{(b) Self-evolving adaptation evaluation} maintains persistent adaptive
state $z_t$, which is updated from adaptation data and carried across stages,
and evaluates whether accumulated experience supports adaptation to newly
introduced capabilities while preserving previously accessible competence.
}
\label{fig:evaluation_protocol}
\end{figure}

\section{Skill Construction Details}
\label{sec:skill_construction_details}

\begin{figure}[t]
\centering
\includegraphics[width=\columnwidth]{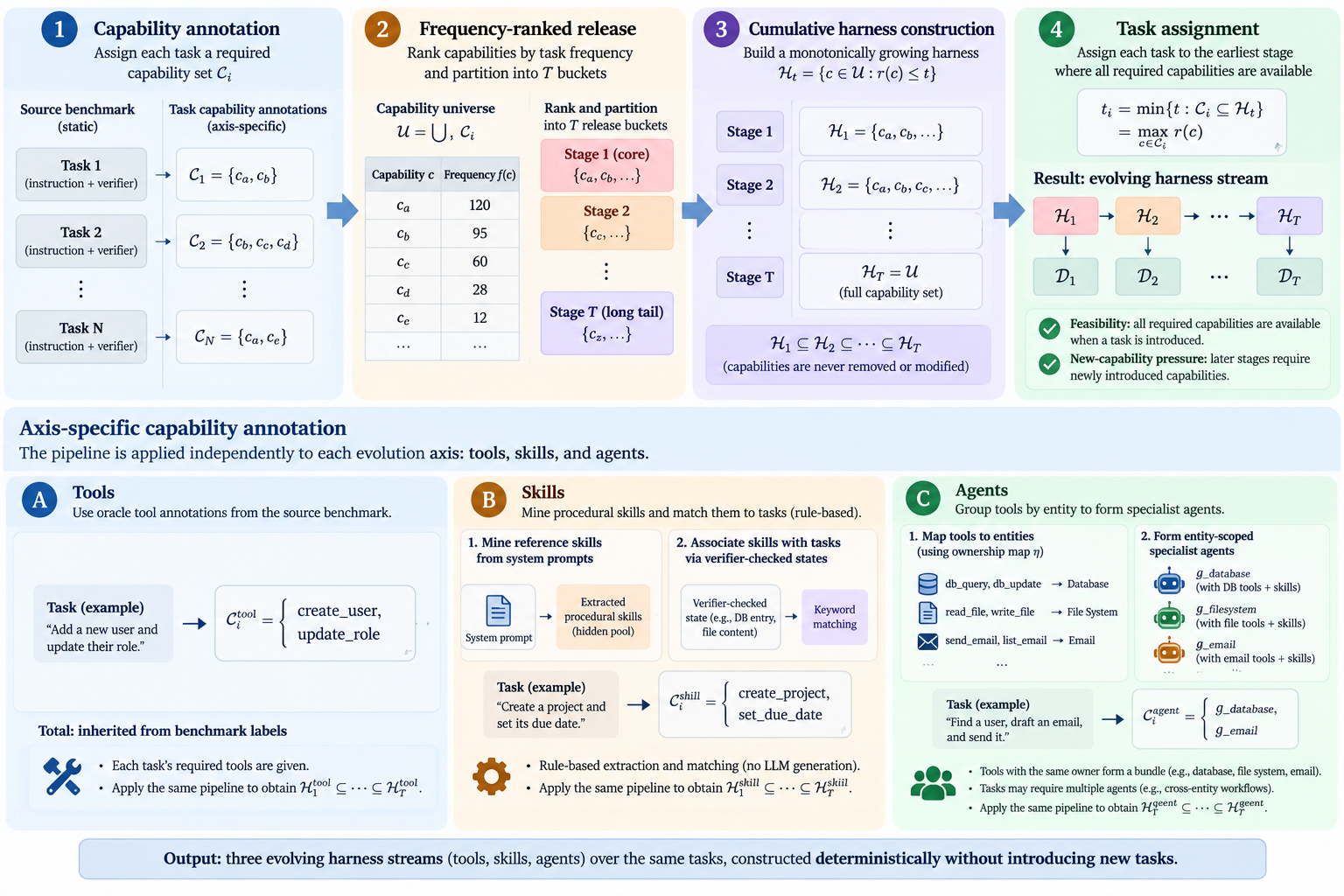}
\vspace{-1\baselineskip}
\captionof{figure}{
Construction pipeline of \ourbenchmark.
\textbf{Top:} Given task-level capability annotations, capabilities are ranked by task frequency, partitioned into stage-wise release buckets, and accumulated into a monotonic harness sequence $\mathcal{H}_1 \subseteq \cdots \subseteq \mathcal{H}_T$. Each task is assigned to the earliest stage at which all of its required capabilities are available, ensuring both feasibility and new-capability pressure.
\textbf{Bottom:} Capability annotations are constructed independently for each evolution axis: tools inherit oracle annotations from the source benchmark, skills are mined and matched to tasks using rule-based procedures, and agents are formed by grouping tools into entity-scoped specialists.
The same deterministic pipeline is then applied separately to obtain evolving tool, skill, and agent harness streams without introducing new tasks.
}
\label{fig:construction_pipeline}
\end{figure}

As described in the main text, we construct task-level skill annotations in two
steps: first mining a deterministic reference-skill library, and then associating
tasks with the reference skills related to their verifier-checked outcomes.
Because EOG and ALE expose procedural information differently, the concrete
implementation differs across the two benchmarks. No generative model is used
in either construction.

\subsection{Mining Reference Skills}

\textbf{EOG.}\quad
EOG domains contain shared system prompts that mix general agent instructions
with domain-specific procedural policies. We parse their section structure and
retain general instructions---such as role definitions, behavioral constraints,
and output requirements---in the agent-facing prompt, while extracting reusable
procedural sections and subsections as reference skills. The extracted procedural
content is removed from the prompt used in the skill track, so evaluated systems
must access it through the provided skill library rather than through
prompt-baked instructions. Domains whose prompts do not contain a sufficiently
decomposable procedural policy are not included in the EOG skill track.

\textbf{ALE.}\quad
ALE does not contain an analogous shared procedural policy that can be separated
from the task prompt. We therefore deterministically define a small catalog of
reusable procedural atoms from recurring workflow and software requirements in
the benchmark. Each atom consists of a canonical procedure together with fixed
lexical patterns and, where applicable, software anchors. Cross-cutting
instructions that apply to nearly all tasks remain visible as baseline
instructions; the remaining atoms form the hidden reference-skill library.
Unlike EOG, no task-specific content is removed from ALE prompts.

\subsection{Associating Skills with Tasks}

\textbf{EOG.}\quad
For each task, we parse its SQL verifiers to recover the table, column, and value
constraints that determine successful execution, and extract the same identifiers
from each reference skill. A skill is associated with a task when the
verifier-checked state provides sufficient matching evidence: a value-level
match is sufficient, while column-level evidence requires multiple matching
constraints. System-managed fields such as identifiers and bookkeeping columns
are ignored. The resulting matched set defines
$\mathcal{C}_i^{\mathrm{skill}}$.

\textbf{ALE.}\quad
Because ALE uses heterogeneous Python scorers rather than a common structured
verifier language, association is lexical. A procedural atom is associated with
a task when its fixed patterns match the task specification or when its
software anchors overlap with the task's normalized required software. The
resulting set again defines $\mathcal{C}_i^{\mathrm{skill}}$.

\textbf{Filtering and releases.}\quad
Reference skills associated with no tasks and tasks with
$\mathcal{C}_i^{\mathrm{skill}}=\varnothing$ are excluded from the skill track.
We retain only domains that support at least three non-empty harness stages.
The remaining reference skills and task annotations are then passed to the
shared frequency-ranked release construction in \cref{sec:construction} to form
$\mathcal{H}_1^{\mathrm{skill}} \subseteq \cdots \subseteq
\mathcal{H}_T^{\mathrm{skill}}$.

\section{Agent Construction Details}
\label{sec:agent_construction_detail}

The agent axis deterministically derives specialist agents from the capability
structure already available in the source benchmarks. As described in the main
text, we first map each tool-like capability to an owner and then convert each
owner bundle into one specialist agent. No LLM or additional human annotation is
used in this construction.

\textbf{Constructing the ownership map.}\quad
For EOG, tools operate on named entities in the underlying enterprise state
(e.g., users, cases, incidents, or knowledge records). We recover the owner from
the tool interface using a deterministic entity lookup, with more specific entity
names matched before more general ones. Tools assigned to the same entity form
one bundle. Thus, each EOG specialist corresponds to an entity and receives all
tools associated with that entity.

ALE does not expose structured function tools; instead, tasks operate through
software available in the execution environment. We therefore apply the same
construction to the canonicalized software units produced by the tool axis,
grouping related software into deterministic software families (e.g., Python
libraries, command-line utilities, or domain-specific software stacks). Each
family becomes one agent owner. In both benchmarks, every capability is assigned
to exactly one owner, yielding a disjoint partition of the available capability
universe.

\textbf{Constructing specialist agents.}\quad
For each owner $e$, we instantiate a specialist agent $g_e$ with (i) a name and
routing description derived from the owner, (ii) access restricted to the
owner's tool or software bundle, and (iii) procedural context associated with
that owner. The lead agent itself has no direct access to environment tools, so
environment interaction must occur through the selected specialists. This
ensures that when a task's required capabilities span multiple owners, successful
execution requires delegation across the corresponding agents.

The procedural context is derived from the same deterministic skill-mining
procedure used for the skill axis. For EOG, mined procedural content is
associated with the entities to which the corresponding operations apply and
attached to the relevant specialists. For ALE, software-anchored procedures are
attached to the corresponding software families, while cross-cutting execution
instructions are shared across agents. No new procedural content is generated
for the agent axis.

\textbf{Task annotations and harness evolution.}\quad
Given the resulting ownership map, a task's required agent set is obtained
directly from the owners of its required capabilities, as defined in the main
text. We then treat the constructed specialist agents as the capability universe
and apply the same frequency-ranked release procedure from
\cref{sec:construction} to obtain the cumulative agent pools
$H_1^{\mathrm{agent}} \subseteq \cdots \subseteq H_T^{\mathrm{agent}}$.
Consequently, later stages expose the lead agent to both agents required by the
current task and previously introduced agents that remain in the cumulative
pool.

\section{System Details}

\subsection{Single-agent Systems}
\label{sec:sas_mem_details}
We evaluate several memory configurations to characterize how the content and representation of adaptation experience affect performance under evolving tools. Unless otherwise specified, memories are constructed from the adaptation split and provided to the agent cumulatively at each stage of the evolving environment. We consider both raw trajectory-based memories and compact memory representations that extract reusable information from prior interactions.

\textbf{Raw (Simple Baseline)} stores the complete interaction history from each prior adaptation episode. For every episode, the memory contains the task description, tool calls and their arguments, execution outcomes, and verifier feedback. These trajectories are retained without filtering, compression, or abstraction and are provided to the agent as stage-wise episodic memory. This configuration serves as the primary baseline for direct replay of prior experience.

\textbf{Raw (task-specific tools in adapt. and eval.)} uses the same raw episodic memory as \textbf{Raw}, but restricts the agent's action space to the ground-truth tools required for the current task. Thus, the agent still receives complete prior trajectories, while irrelevant tools are removed from consideration during execution. This configuration separates the effect of experience from the difficulty of selecting among an evolving set of tool candidates.

\textbf{Raw (task-specific tools in adapt. only)} further provides oracle access to the complete test-time tool configuration while retaining the raw adaptation memory. This condition removes both irrelevant tool-selection choices and uncertainty arising from the evolving test-time tool set, providing an upper-bound-style control for evaluating the contribution of memory under favorable tool availability.

\textbf{Raw + Success Only} retains only adaptation episodes that were successfully completed, discarding unsuccessful trajectories and their verifier feedback. The resulting memory therefore consists exclusively of successful demonstrations, while preserving the complete task descriptions and tool-use traces within those episodes. This configuration evaluates the effect of selectively retaining positive adaptation experience.

\textbf{Raw (successful traj. only)} retains only unsuccessful adaptation episodes, including their task descriptions, tool-use traces, execution outcomes, and verifier feedback. This configuration corresponds to \texttt{memory_filter=failure} and evaluates whether unsuccessful experience provides useful information for subsequent adaptation, despite not containing successful execution traces.

\textbf{Reasoning Bank}~\citep{ouyang2026reasoningbank} replaces complete trajectory replay with compact reasoning artifacts extracted from prior adaptation episodes. Each episode is summarized into reusable reasoning items containing a title, description, and procedural reasoning derived from the observed interaction and outcome. Tool-use traces are not retained in their original form. This configuration therefore tests whether abstract procedural knowledge extracted from experience can provide more useful adaptation guidance than replaying complete trajectories.

\textbf{Reasoning Bank (+ tool information)} augments the reasoning artifacts with information about the tools encountered during adaptation. In addition to the abstract reasoning extracted from each episode, the memory retains the relevant tool identities and interface information needed to connect the reasoning to available actions. This condition evaluates whether explicitly preserving tool-related information improves the usefulness of otherwise abstract reasoning memories in evolving tool environments.

\textbf{MemToolAgent}~\citep{er2026memtoolagent} represents each adaptation episode as a reflection rather than as a complete trajectory. The reflection summarizes the agent's experience, including relevant actions, environmental feedback, and task outcome, into concise procedural guidance that can be retrieved for subsequent tasks. This representation emphasizes reusable lessons from prior interactions while discarding execution-specific details that are preserved by raw trajectory memory.

\subsection{Multi-Agent Systems}
\label{sec:mas_tool_detail}

We provide detailed configurations of the multi-agent systems used in our evaluation. Since several methods were originally designed for less agentic settings, we make minor adaptations to fit our evaluation protocol while preserving their core architectures and design principles.

\textbf{AutoGen}~\citep{wu2024autogen}.\quad
Under \textit{evolving tools}, AutoGen uses a single solver agent that directly interacts with the available tools for up to 50 iterations. To mitigate repeated-action stalls, a secondary ground-truth role intervenes when the solver issues the same tool call twice consecutively. Under \textit{evolving agents}, the solver instead acts purely as a router: it selects an appropriate specialist and delegates the task, after which the specialist executes independently in a fresh sub-conversation with a 12-iteration budget.

\textbf{G-Memory}~\citep{zhang2025g-memory}.\quad
G-Memory follows the same execution structure as AutoGen (as its backbone), while augmenting it with persistent memory. Under \textit{evolving tools}, the system stores task-level successes and failures in a hierarchical memory and retrieves relevant prior experience for subsequent tasks. Under \textit{evolving agents}, the memory is adapted to record delegation decisions and their outcomes, thereby supporting specialist selection rather than direct tool use.

\textbf{LegoMem}~\citep{han2025legomemmodularproceduralmemory}.\quad
LegoMem natively adopts an orchestrator-worker architecture. Under \textit{evolving tools}, an orchestrator decomposes each task into 2-5 subproblems and assigns them to worker agents, each with a budget of 25 iterations. Workers share an intermediate workspace and retrieve memories of prior task decompositions and executions. By default, however, memories are maintained separately for each worker rather than shared across workers; cross-worker memory sharing is evaluated only as an ablation variant (\textit{shared memory}) in Figure~\ref{fig:mas-radar}.
Under \textit{evolving agents}, the same decomposition procedure is retained, but each subproblem is assigned directly to a named specialist from the current agent roster.

\textbf{DeLM}~\citep{mao2026decentralized}.\quad
DeLM uses an orchestrator-free, parallel execution scheme in which up to three workers independently process ready subproblems from a shared task pool, with at most three replanning rounds and 12 subproblems per task. 
By default, each agent maintains a raw memory that records its previous behavior, intermediate results, and outcomes.
Under \textit{evolving tools}, newly acquired information is verified against the originating worker's execution before being added to the shared workspace. Under \textit{evolving agents}, each subproblem is additionally routed to a specific specialist, and the shared workspace records the provenance of specialist contributions.

\section{Claude Code Results}
\label{sec:claude_code_result}

Claude Code is included as an additional frontier deployment system for the
skill and agent axes. Because Claude Code differs from the controlled
GPT-5/Codex systems in both its underlying model and native agent harness, we do
not use its absolute performance for direct cross-system comparisons in the main
tables. Its \emph{within-system} comparison between task-specific and cumulative
harnesses nevertheless provides an informative reference for how another
frontier agent responds to harness evolution. We therefore report the complete
aggregate results here and retain representative transfer results in the main
text.

\begin{table}[t]
\caption{\small
Claude Code deployment results under evolving skills and agents. Task-specific
conditions expose only the capabilities annotated for each task, while
cumulative conditions expose the full harness available at the corresponding
stage. These results are intended for within-Claude-Code comparisons rather than
direct comparison with the GPT-5/Codex-based systems in the main tables.}
\vspace{-.8\baselineskip}
\centering
{%
\setlength{\dashlinedash}{0.7pt}%
\setlength{\dashlinegap}{1.2pt}%
\setlength{\arrayrulewidth}{0.2pt}%
\resizebox{\textwidth}{!}{%
\begin{tabular}{
  @{\hskip 4pt}ll@{\hskip 6pt}
  cccc@{\hskip 7pt}:
  cccc@{\hskip 6pt}:
  c@{\hskip 4pt}
}
\toprule
& &
\multicolumn{4}{c}{{\large\textsf{\textbf{\ourbenchmark-EOG}}}}
& \multicolumn{4}{c}{{\large\textsf{\textbf{\ourbenchmark-ALE}}}}
& \textbf{Overall} \\
\textbf{Axis} & \textbf{Harness}
& Pass (\%) & Score & \faIcon[regular]{clock} (h) & Tok. (M)
& Pass (\%) & Score & \faIcon[regular]{clock} (h) & Tok. (M)
& \textbf{Pass Rate} \\
\midrule

\multirow{2}{*}{Skills}
& Task-specific
& 29.3$^{\pm0.8}$ & 65.5$^{\pm0.5}$ & 2.5 & 19.1
& 13.2$^{\pm0.0}$ & 35.0$^{\pm0.3}$ & 36.0 & 389.7
& 24.2 \\
& Cumulative
& 30.2$^{\pm1.4}$ & 67.3$^{\pm1.0}$ & 3.8 & 13.6
& 13.2$^{\pm1.2}$ & 35.4$^{\pm1.5}$ & 36.7 & 422.7
& 24.8 \\
\midrule

\multirow{2}{*}{Agents}
& Task-specific
& 11.0$^{\pm1.7}$ & 47.2$^{\pm0.7}$ & 10.3 & 7.1
& 16.9$^{\pm2.0}$ & 41.3$^{\pm1.4}$ & 32.4 & 240.3
& 12.8 \\
& Cumulative
& 10.6$^{\pm1.1}$ & 49.5$^{\pm0.3}$ & 10.8 & 7.0
& 15.3$^{\pm3.3}$ & 42.3$^{\pm3.1}$ & 33.1 & 255.2
& 12.0 \\

\bottomrule
\end{tabular}}%
\par}
\vspace{-1\baselineskip}
\label{tab:claude_code_results}
\end{table}

The within-system trends are consistent with the qualitative deployment findings
in the main paper. Under evolving skills, cumulative exposure has little aggregate
effect: Claude Code's EOG pass rate changes from $29.3\%$ to $30.2\%$, while
ALE remains at $13.2\%$. Under evolving agents, cumulative exposure has a modest
negative effect, with pass rate changing from $11.0\%$ to $10.6\%$ on EOG and
from $16.9\%$ to $15.3\%$ on ALE. These results provide an additional frontier-system
reference for the conclusion that the effect of harness expansion is axis-dependent,
while avoiding uncontrolled absolute comparisons with the GPT-5/Codex systems.

\section{Additional Analysis on Evolving Tools}
\label{sec:evovle_tool_details}

\subsection{Single-Agent Systems}
\begin{wraptable}{r}{0.58\columnwidth}
\vspace{-4\baselineskip}
\caption{\small Memory ablations for SAS under evolving tools (\ourbenchmark-EOG). We compare raw memory with variants that selectively retain successful or failed experiences, incorporate oracle tools, or use alternative memory mechanisms. 
}
\vspace{-1\baselineskip}
\centering
{%
\setlength{\dashlinedash}{0.7pt}%
\setlength{\dashlinegap}{1.2pt}%
\setlength{\arrayrulewidth}{0.2pt}%
\resizebox{0.55\textwidth}{!}{%
\begin{tabular}{
  @{\hskip 4pt}l@{\hskip 8pt}
  cc
}
\toprule
& Pass (\%) & Score \\
\midrule
\textbf{Raw} & \hscb{33.0}$^{\pm0.9}$ & \hscb{66.2}$^{\pm0.6}$ \\
$\;$ task-specific tools in adapt. and eval. & \hscb{33.3}$^{\pm0.4}$ & \hscb{66.6}$^{\pm0.6}$ \\
$\;$ task-specific tools in adapt. only & \hscb{35.3}$^{\pm0.4}$ & \hscb{67.7}$^{\pm0.1}$ \\
$\;$ successful traj. only & \hscb{34.4}$^{\pm1.7}$ & \hscb{65.3}$^{\pm0.8}$  \\
$\;$ failed traj. only & \hscb{34.4}$^{\pm0.7}$ & \hscb{66.1}$^{\pm0.5}$  \\
\midrule
\textbf{Reasoning Bank}~\citep{ouyang2026reasoningbank} & \hscb{36.9}$^{\pm1.3}$ & \hscb{68.7}$^{\pm0.4}$ \\
$\;+$ tool information & \hscb{30.3}$^{\pm1.0}$ & \hscb{63.4}$^{\pm0.5}$ \\
\textbf{MemToolAgent}~\citep{er2026memtoolagent} & \hscb{38.6}$^{\pm1.0}$ & \hscb{68.9}$^{\pm0.5}$  \\
\bottomrule
\end{tabular}}%
\par}
\label{tab:sas_tool_ablation}
\vspace{-1\baselineskip}
\end{wraptable}
In this section, we provide a breakdown analysis of memory-based adaptation under the \textit{evolving tools} setting. Table~\ref{tab:sas_tool_ablation} compares memory variants that differ in the representation and content of retained experience, while Table~\ref{tab:tool_adaptation_harness_raw} examines how the adaptation harness itself affects the usefulness of accumulated memory.

The main results show that memory-based self-evolution accounts for several of the strongest gains under evolving tools. We therefore isolate how the \emph{form and content of retained experience} affect adaptation. Rather than assuming that more memory is always beneficial, we ask which properties of prior experience provide useful signals as the tool catalog expands.

We ablate memory along three dimensions: \emph{representation} (raw trajectories vs.\ structured reasoning or reflection), \emph{content} (tool information and trajectory outcomes).

\textbf{Structured experience is more effective than raw replay.}\quad
Memory-based self-evolution benefits from transforming prior experience into structured representations. Raw trajectory memory improves pass rate from $30.2\%$ to $33.0\%$, while Reasoning Bank and MemToolAgent achieve $36.9\%$ and $38.6\%$, respectively. This suggests that the benefit of accumulated experience depends on how it is represented, rather than simply on retaining additional history.

\textbf{More tool information is not necessarily better.}\quad
Adding explicit tool information to Reasoning Bank substantially reduces performance from $36.9\%$ to $30.3\%$. Thus, under an evolving tool catalog, exposing additional interface details within memory can introduce interference rather than improve adaptation. Other ablations on tool identity and trajectory outcomes show smaller differences.

As shown in \cref{tab:tool_adaptation_harness}, Raw Memory adapted under the
full cumulative tool catalog achieves $33.0\%$ pass rate and $66.2$ score on
EOG. When adaptation instead uses task-specific tools, while evaluation remains
under the same cumulative catalog, pass rate increases to $35.3\%$, score to
$67.7$, and adaptation time decreases from $25.7$ to $20.5$ hours. Because the
evaluation harness is identical, this difference points to the adaptation
context itself: experience grounded in task-relevant tools can produce more
useful persistent state than experience acquired while reasoning over the
broader cumulative tool space.

\textbf{What Makes Past Experience Useful as the Tool Harness Evolves?}\quad

As shown in \cref{tab:tool_main_results}, memory-based self-evolving adaptation accounts for several of the strongest gains under evolving tools, with ReasoningBank and MemToolAgent among the best-performing methods. We therefore use memory as a concrete lens to study how the content and representation of past experience affect performance as the tool harness evolves.
We ablate $z_t$ along three dimensions: \emph{content} (what is stored), \emph{valence} (success vs.\ failure), and \emph{fidelity} (real vs.\ corrupted tool identifiers). Our memory ablations (detailed in 
\cref{tab:sas_tool_ablation}) show that
\textbf{how experience is represented matters more than simply retaining more
experience}. 
\begin{wraptable}{r}{0.5\columnwidth}
\caption{\small Effect of the adaptation harness for Raw Memory on EOG. All conditions follow the same task stream and are evaluated under the cumulative tool harness; ``$-$ broad exposure'' uses only task-specific tool during
adaptation;
}
\vspace{-1.0em}
\centering
\small
\setlength{\tabcolsep}{4pt}
\resizebox{\linewidth}{!}{%
\begin{tabular}{l l c c c}
\toprule
\textbf{Adapt.} & \textbf{Eval.} &
\textbf{Pass} & \textbf{Score} & \textbf{Time} \\
\textbf{Harness} & \textbf{Harness} &
\textbf{(\%)} & & \textbf{(h)} \\
\midrule
Cumulative    & Cumulative & 33.0 & 66.2 & 25.7 \\
$\;\;-\;$ broad exposure & Cumulative & \textbf{35.3} & \textbf{67.7} & \textbf{20.5} \\
\bottomrule
\end{tabular}
}
\label{tab:tool_adaptation_harness_raw}
\end{wraptable}
Structured memory consistently outperforms raw trajectory replay:
MemToolAgent achieves $38.6\%$ pass rate compared with $33.0\%$ for raw memory
and $30.2\%$ without memory. Moreover, adding explicit tool schemas to
Reasoning Bank reduces performance from $36.9\%$ to $30.3\%$, suggesting that
additional tool information can introduce interference rather than improve
adaptation. 
Finally, success-only and failure-only memories perform similarly
($34.4\%$), indicating that unsuccessful interactions can provide useful
signals for adapting to an evolving tool space. Together, these results
suggest that effective self-evolving adaptation depends on \textbf{selectively
extracting actionable experience}, rather than simply accumulating more
context.

\subsection{Multi-Agent Systems}
\label{app:mas_tool_results}

\begin{wrapfigure}{r}{0.4\columnwidth}
\centering
\includegraphics[width=0.4\columnwidth]{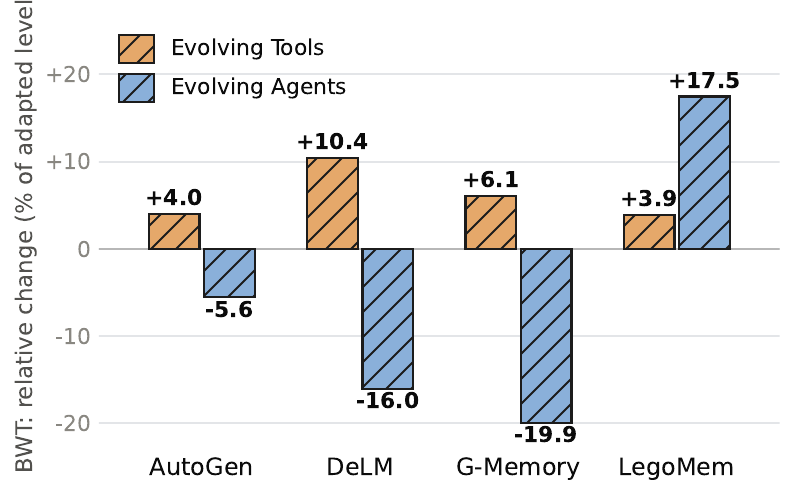}
\captionof{figure}{\small Average Relative Backward Transfer (BWT) on \ourbenchmark-EOG of MAS. 
}
\vspace{-2em}
\label{fig:mas-bwt}
\end{wrapfigure}
\begin{figure}[t]
\centering
\includegraphics[width=.9\columnwidth]{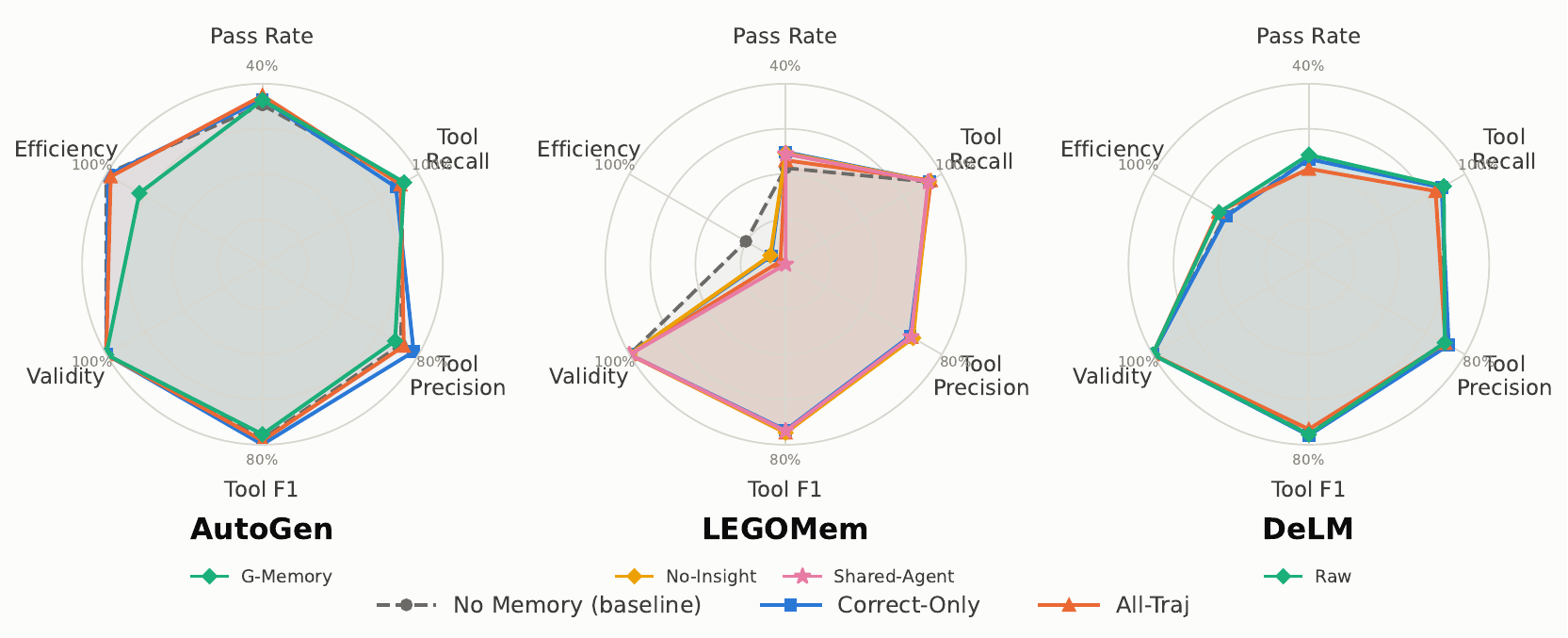}
\vspace{-1\baselineskip}
\captionof{figure}{
Ablation study of MAS memory behaviors on \ourbenchmark-EOG. We focus on two commonly used components of agentic memory systems: (i) high-level insights abstracted from execution trajectories, and (ii) the choice of whether to retain only successful trajectories or to also incorporate failed trajectories into memory.
}
\label{fig:mas-radar}
\end{figure}

In this section, we provide a breakdown analysis of multi-agent systems and their memory variants under the \textit{evolving tools} setting. 
Figure~\ref{fig:mas-bwt} reports relative backward transfer (BWT) under both the \textit{evolving tools} and \textit{evolving agents} settings, while Figure~\ref{fig:mas-radar} compares different MAS memory variants under the \textit{evolving tools} setting.

\textbf{MAS exhibits little forgetting under evolving tools, but persistent memory provides no clear retention advantage over a no-memory baseline.}
For MAS, BWT remains uniformly small in magnitude, with all methods falling within $\pm 2.3$ percentage points. The no-memory AutoGen baseline, DeLM, and G-Memory show mild positive BWT ($+2.3\%$, $+1.2\%$, and $+0.4\%$, respectively), while LegoMem is the only method with negative transfer ($-1.7\%$). This suggests that, overall, performance on earlier task cohorts is largely preserved as the tool harness evolves. Moreover, LegoMem's degradation is concentrated in only two domains rather than broadly distributed, pointing to a mechanism-specific failure mode rather than a general limitation of memory-augmented adaptation.

Notably, however, none of the memory-based methods improves BWT over the no-memory AutoGen baseline, which achieves the highest value ($+2.3\%$). Thus, unlike the SAS setting, persistent memory does not provide a clear retention advantage for MAS under evolving tools. This comparison should nevertheless be interpreted cautiously: high BWT alone does not imply effective continual adaptation, since a weaker baseline may appear stable simply because it acquires less new competence and therefore has less useful knowledge to forget. Overall, the small spread in BWT suggests that retention is not the primary differentiator among MAS memory mechanisms; their effectiveness must instead be assessed jointly with their ability to adapt to newly introduced tasks.

\paragraph{Memory can constrain or amplify search in multi-agent systems.}

Across \ourbenchmark-EOG, memory exhibits strongly architecture-dependent effects under an evolving tool harness. \textbf{Most notably, adding memory does not consistently improve adaptation: its benefit depends more on what is stored and how it is used than on the mere presence of memory.} In AutoGen, concrete trajectory memory primarily sharpens tool selection: storing prior trajectories improves pass rate while reducing tool calls and failed verifications, and successful-only trajectories achieve the highest tool precision (77.4\%) and exact-match rate (15.8\%). In contrast, G-Memory's abstract summaries \textit{encourage broader exploration, increasing calls and redundancy while lowering precision, }suggesting that abstraction alone does not effectively constrain an expanding tool space. LEGOMem shows the opposite pattern: its gains come not from more precise tool selection, but from broader execution and fewer failed verifications; removing high-level insights has little effect, whereas incorporating failed trajectories weakens the gain and sharing memory across sub-agents substantially increases execution cost. \textbf{DeLM provides the clearest counterexample to the assumption that richer memory is always better:} its default memory lowers pass rate by 2.2 points, while removing incorrect trajectories restores baseline performance and removing distillation yields the best result (+0.8 points). \textbf{Taken together, the strongest recurring signal is that concrete successful trajectories are consistently safer and more useful than abstracted, failed, or uniformly shared memories across sub-agents.}
Thus, under harness evolution, multi-agent memory should be viewed not as a uniformly beneficial module, but as an architecture-sensitive mechanism whose effectiveness depends on memory quality, granularity, and placement.

\textbf{SAS vs. MAS: Experience benefits depend on agent architecture.} 
Self-evolution is architecture-dependent: accumulating experience consistently helps SAS, but its benefits become less predictable in MAS, suggesting that experience must be aligned not only with the task and harness, but also with how agents coordinate. In SAS, memory-based methods consistently improve over the deployment baseline, with MemToolAgent achieving the strongest overall performance ($35.1\%$). In contrast, multi-agent systems show substantially more variable gains: G-Memory reaches $30.1\%$, while LegoMem achieves only $20.6\%$. These results suggest that experience that is effective when directly integrated by a single agent does not necessarily translate to multi-agent settings, where experience must be coordinated across agents and may interact with their individual decision processes.

\section{Additional Analysis on Evolving Skills}
\label{sec:evovle_skill_details}

\subsection{Skill invocation details}

\begin{wraptable}{r}{0.58\columnwidth}
\vspace{-1.3em}
\caption{\small Memory-based ablations under evolving skills (\ourbenchmark-EOG).  
}
\vspace{-.8\baselineskip}
\centering
{%
\setlength{\dashlinedash}{0.7pt}%
\setlength{\dashlinegap}{1.2pt}%
\setlength{\arrayrulewidth}{0.2pt}%
\resizebox{0.55\textwidth}{!}{%
\begin{tabular}{
  @{\hskip 4pt}l@{\hskip 8pt}
  cc
}
\toprule
& Pass (\%) & Score  \\
\midrule
Raw & \hscb{17.7}$^{\pm6.8}$ & \hscb{55.2}$^{\pm4.4}$ \\
$\;$ task-specific tools in adapt. only & \hscb{23.9}$^{\pm1.4}$ & \hscb{60.4}$^{\pm0.3}$\\
Fake-tool memory & \hscb{17.5}$^{\pm4.8}$ & \hscb{54.7}$^{\pm7.1}$ \\
Reasoning-playbook memory & \hscb{16.5}$^{\pm5.6}$ & \hscb{52.2}$^{\pm6.2}$ \\
MemToolAgent~\citep{er2026memtoolagent} & \hscb{4.4}$^{\pm1.9}$ & \hscb{19.7}$^{\pm2.2}$ \\
\bottomrule
\end{tabular}}%
}
\vspace{-1\baselineskip}
\label{tab:skill_memory_ablation}
\end{wraptable}

\textbf{Tool-oriented experience does not transfer cleanly to skill adaptation.}
Given that skill engagement is a central bottleneck above, we next ask whether memory mechanisms that are effective for evolving tools can help. Here, task-relevant skills are already provided, while the retained memories mainly encode tool-use and action experience rather than when or how to engage skills; accordingly, most memory variants remain close to the no-memory baseline. MemToolAgent is a stronger failure case: despite being the best memory method under evolving tools, its pass rate drops to $4.4\%$ (\cref{tab:skill_memory_ablation}). This sharp degradation suggests negative transfer from tool-centric experience, potentially because the learned reflections bias execution toward tool-use patterns that are poorly aligned with the provided procedural skills. Together,
these results suggest that persistent experience must be aligned with the type of
capability being evolved.

\textbf{Skill invocation details.}
\textbf{(a)}~Share of each system's offered skill library that is actually invoked
(a distinct \texttt{SKILL.md} opened).
\textbf{(b)}~Precision and \textbf{(c)}~recall of those invoked skills against the
task-specific skill set.
Bars are last-stage only (CSM $v_3$, HR $v_3$, ITSM $v_4$), pooled task-weighted over
the three domains.
GPT-5 unless labeled otherwise.
Each bar uses that arm's own library as the denominator: the two task-specific Codex
controls mount one task's skills, while Task-specific GEPA and the cumulative arms
mount the stage library.
\begin{wrapfigure}{r}{0.5\columnwidth}
\centering
\vspace{-1em}
\includegraphics[width=\linewidth]{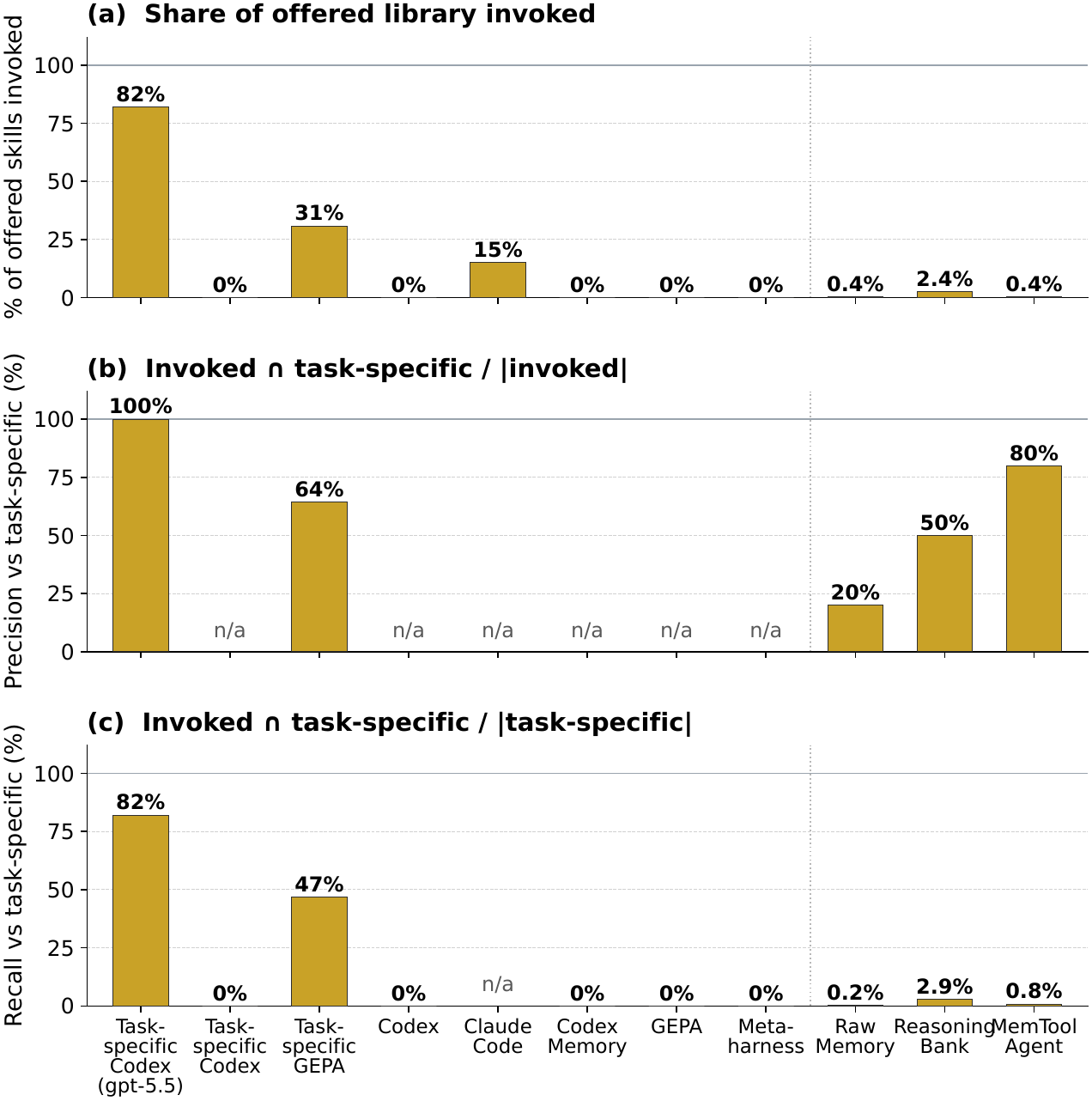}
\vspace{-1.6\baselineskip}
\captionof{figure}{\small Skill invocation and overlap with task-specific skills at the final stage.}
\label{fig:skill_invocation_share1}
\vspace{-2\baselineskip}
\end{wrapfigure}
Precision is omitted (\textit{n/a}) when nothing is invoked.
Claude Code's share counts file reads rather than distinct skills, and its
precision/recall are omitted because skill names were not logged.

\subsection{Skill learning methods}
\label{app:skill_learn_details}
We describe the skill-learning methods that we evaluate as methods for adapting to an expanding skill-harness below:

\textbf{Zero-shot.}\quad
For each test task, the curator constructs a private skill overlay using only the task's visible instruction, its listed gold tool names, and the library snapshot available at that stage. The curator has no access to training tasks, training trajectories, or verifier outcomes. Thus, the resulting skills are conditioned on the current test task rather than distilled from prior solved experience.

\textbf{Raw Traj.}\quad
The skill library is initialized empty at $h_1$ and evolves throughout the curriculum. After the training trials at stage $h_k$, a batch curator processes the agents' compact trajectories—including messages, tool calls, and outputs—together with verifier pass/fail outcomes, and uses them to update a shared skill library. The resulting library is reused for test tasks at $h_k$ and carried forward to $h_{k+1}$. Unlike the zero-shot setting, skills are derived directly from agents' execution experience rather than from gold skills or teacher-provided guidance.

\textbf{Empty skills.}\quad
This corresponds to an empty skill directory.

The other skill learning methods: One-shot, Self-feedback, Batch Self-feedback, Batch Teacher feedback and Skill-creator are from \citep{zhong2026skilllearnbenchbenchmarkingcontinuallearning}.

\begin{table}[h]
\caption{\small Skill-learning ablations (Codex with GPT-5 on \ourbenchmark-EOG). Methods vary in how skills are generated and refined from adaptation experience.}
\vspace{-.8\baselineskip}
\centering
{
\setlength{\dashlinedash}{0.7pt}
\setlength{\dashlinegap}{1.2pt}
\setlength{\arrayrulewidth}{0.2pt}
\resizebox{0.6\textwidth}{!}{%
\begin{tabular}{
  @{\hskip 4pt}l@{\hskip 8pt}
  cc
}
\toprule
& Pass (\%) & Score  \\
\midrule
Empty skills & \hscb{16.9}$^{\pm2.5}$ & \hscb{56.1}$^{\pm1.4}$ \\
Zero-shot & \hscb{17.6}$^{\pm1.7}$ & \hscb{54.5}$^{\pm1.3}$ \\
One-shot~\citep{zhong2026skilllearnbenchbenchmarkingcontinuallearning} & \hscb{11.5}$^{\pm0.6}$ & \hscb{41.1}$^{\pm0.4}$ \\
Raw traj.& \hscb{20.0}$^{\pm1.7}$ & \hscb{59.3}$^{\pm0.6}$ \\
Self Feedback~\citep{zhong2026skilllearnbenchbenchmarkingcontinuallearning} & \hscb{12.8}$^{\pm2.2}$ & \hscb{42.1}$^{\pm0.8}$ \\
Batch Self Feedback~\citep{zhong2026skilllearnbenchbenchmarkingcontinuallearning} & \hscb{16.2}$^{\pm2.4}$ & \hscb{57.3}$^{\pm2.5}$ \\
Batch Teacher Feedback~\citep{zhong2026skilllearnbenchbenchmarkingcontinuallearning} & \hscb{22.1}$^{\pm3.0}$ & \hscb{59.7}$^{\pm0.8}$ \\
Skill Creator~\citep{zhong2026skilllearnbenchbenchmarkingcontinuallearning} & \hscb{20.9}$^{\pm1.5}$ & \hscb{60.6}$^{\pm1.0}$ \\
\bottomrule
\end{tabular}}%
\par}

\label{tab:skill_learning_ablation}
\end{table}

\begin{figure}[h]
\centering
\includegraphics[width=0.7\linewidth]{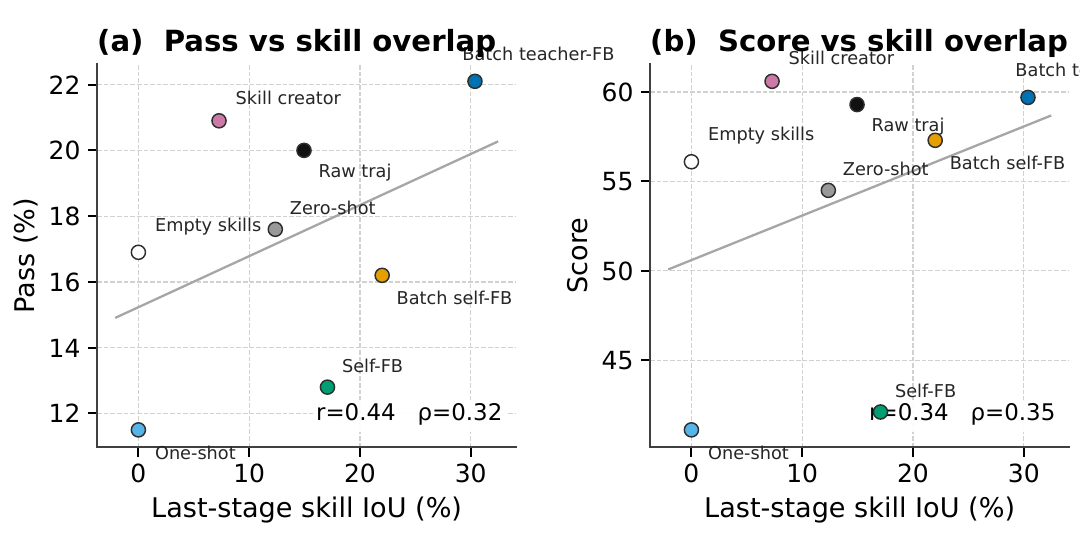}
\vspace{-1\baselineskip}
\captionof{figure}{\textbf{Relationship between last-stage skill overlap (IoU) and downstream performance} on EOG (Codex, GPT-5). Each point corresponds to a skill-learning method; IoU is averaged over CSM~$h_3$, HR~$h_3$, and ITSM~$h_4$, while Pass and Score are taken from the final row of Table~X. Error bars denote s.d. across domains for IoU and the reported s.d. for Pass/Score. $r$ and $\rho$ denote Pearson and Spearman correlations, respectively ($n=7$); 
}
\label{fig:skill_overlap_vs_performance}
\vspace{-1\baselineskip}
\end{figure}

\paragraph{Skill learning: feedback quality determines skill quality.}
\begin{itemize}[leftmargin=*,noitemsep]
    \item \textbf{Execution experience is necessary.} Skills extracted from raw trajectories ($20.0\%$) substantially outperform both empty skills ($16.9\%$) and test-conditioned zero-shot generation ($17.6\%$). Without observing how tasks are actually solved, proposed skills lack grounding.
    \item \textbf{Iterative refinement with feedback matters most.} Batch Teacher Feedback ($22.1\%$), Skill Creator ($20.9\%$)
    are the strongest methods---all use some form of iterative refinement where skills are updated based on execution outcomes.
    \item \textbf{Naive skill learning fails.} One-shot generation ($11.5\%$) and Self Feedback ($12.8\%$) perform \emph{worse} than having no skills at all ($16.9\%$). Poorly generated skills actively mislead the agent---they are worse than nothing.
    \item \textbf{Batch processing helps.} Batch Self Feedback ($16.2\%$) substantially outperforms single-episode Self Feedback ($12.8\%$), suggesting that aggregating experience across multiple tasks before refining skills produces more robust procedures.
    \item {\bf Overlap vs performance.} \Cref{fig:skill_overlap_vs_performance} shows that overlap with task-specific skills is only weakly associated with task performance ($r \approx 0.44$ for Pass and $r \approx 0.34$ for Score). In other words, closer alignment with the gold skill library does not consistently translate into better task performance, indicating that skill overlap captures only part of what matters for effective adaptation.

\end{itemize}

\begin{figure}[h]
\centering
\includegraphics[width=0.7\linewidth]{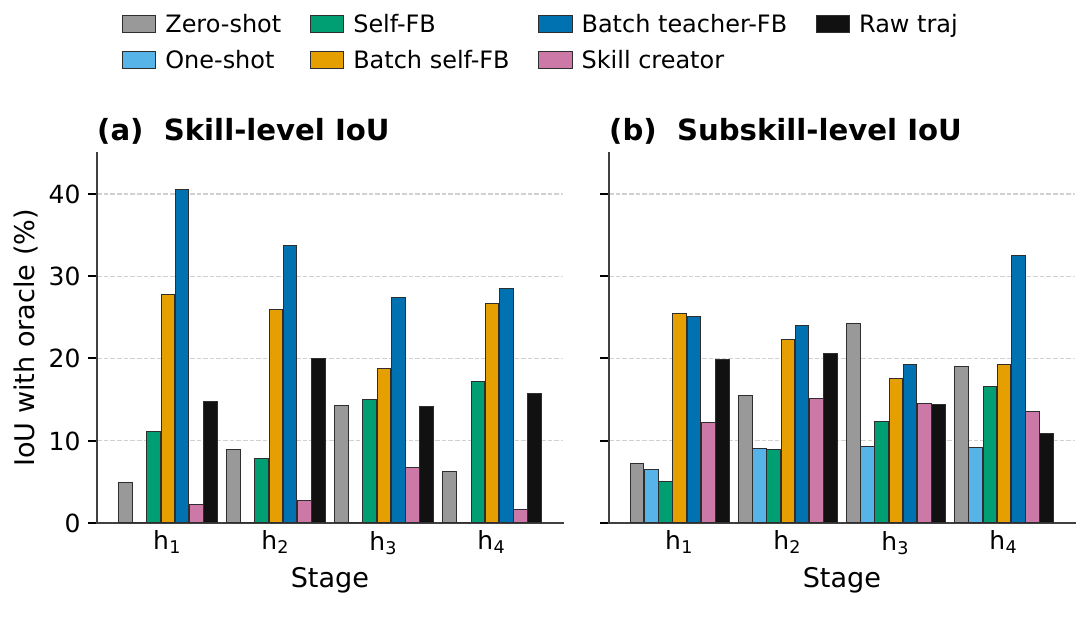}
\vspace{-1.6\baselineskip}
\captionof{figure}{\small   Overlap of generated skills with the task-specific library across the curriculum. This is compute at stage-level since the skills are learned at every stage of the evolving-harness setup. The task-specific skills and sub-skills of all tasks in a stage are compared against the respective learned skills to get the iou.}
\label{fig:skill_iou}
\vspace{-1\baselineskip}
\end{figure}

\paragraph{Skill invocation is a model/agent property.}
At the final curriculum stage, skill invocation depends more on the underlying model and agent than on simply providing a skill library or memory store (\cref{fig:skill_invocation_share1}). GPT-5-based Codex variants leave the offered skills entirely unread, while GPT-5.5 Codex invokes $82\%$ of its task-specific skills. Task-specific GEPA also invokes skills ($31\%$ of those offered), though with lower precision ($64\%$). Memory-based methods remain largely inactive, invoking only $0.4$--$2.4\%$ of the cumulative library. Thus, \textbf{providing or storing skills does not ensure their use: effective skill invocation is primarily an agent/model behavior.}
.
\section{Additional Analysis on Evolving Agents}
\label{sec:evovle_agent_details}

\begin{figure}[h]
\centering
\includegraphics[width=0.5\columnwidth]{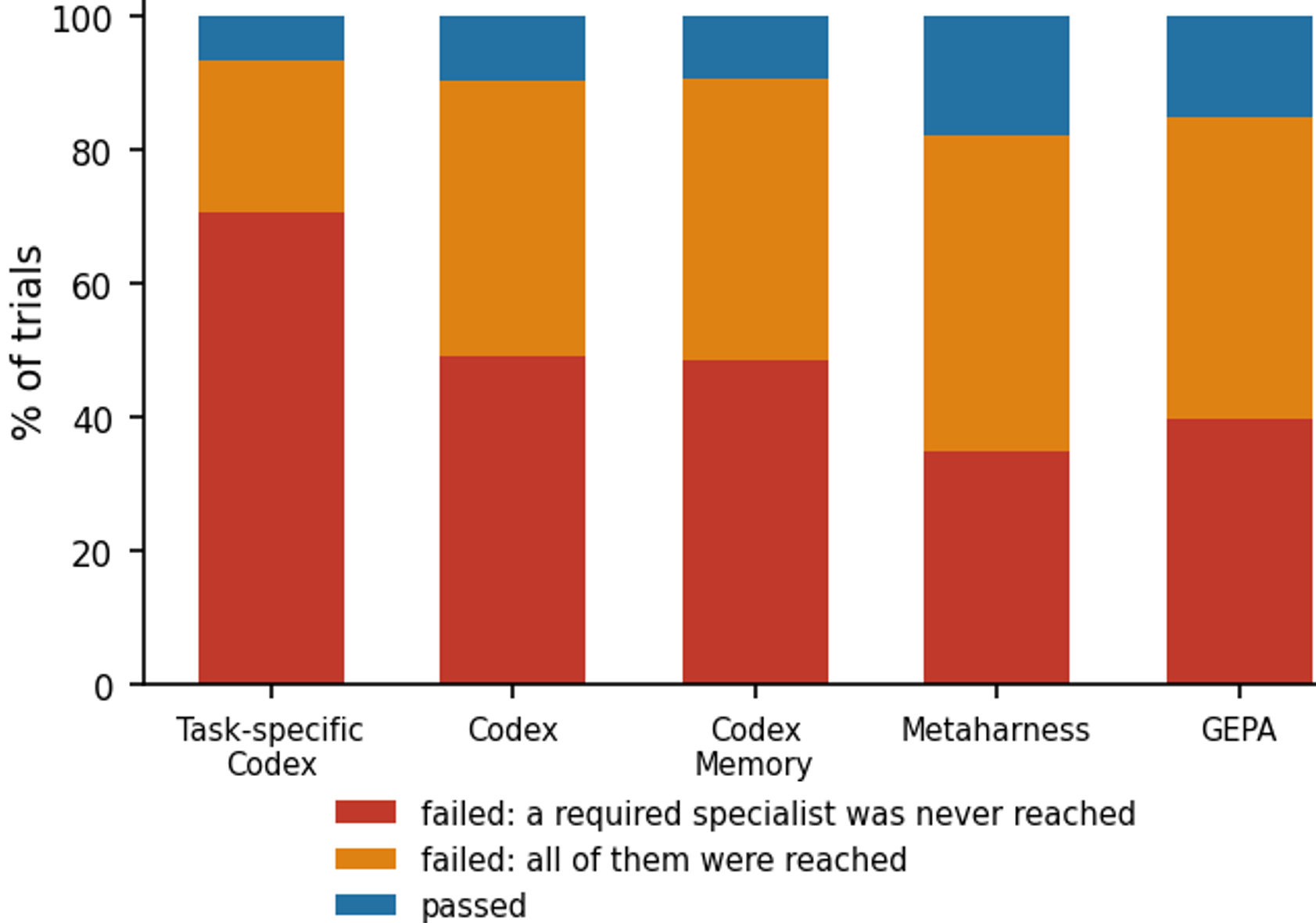}
\vspace{-.5\baselineskip}
\captionof{figure}{\small Failure decomposition.
Trials are divided into successes, failures that miss at least one required agent, and failures despite reaching all required agents.
}
\label{fig:agent_routing_failures}
\vspace{-1\baselineskip}
\end{figure}

\paragraph{Correct delegation is necessary but far from sufficient.}
\Cref{fig:agent_routing_failures} separates failures into those where at least one
required specialist is never reached and those where all required specialists are
reached but the task still fails. Self-evolution clearly reduces the first type:
routing failures account for 54\% of failures in the non-adaptive system, compared
with 47\% for GEPA and 43\% for Meta-Harness. Yet the remaining coordination
problem is substantial. Even conditional on reaching every required specialist,
78--86\% of trials still fail their verifier. Thus, reaching the correct team is
important but insufficient. The agent axis contains two distinct bottlenecks:
\emph{delegation}, identifying and reaching the relevant specialists under an
expanding pool, and \emph{coordination/execution}, successfully combining their
work once the right specialists have been reached.

\begin{figure}[h]
\centering
\includegraphics[width=0.5\columnwidth]{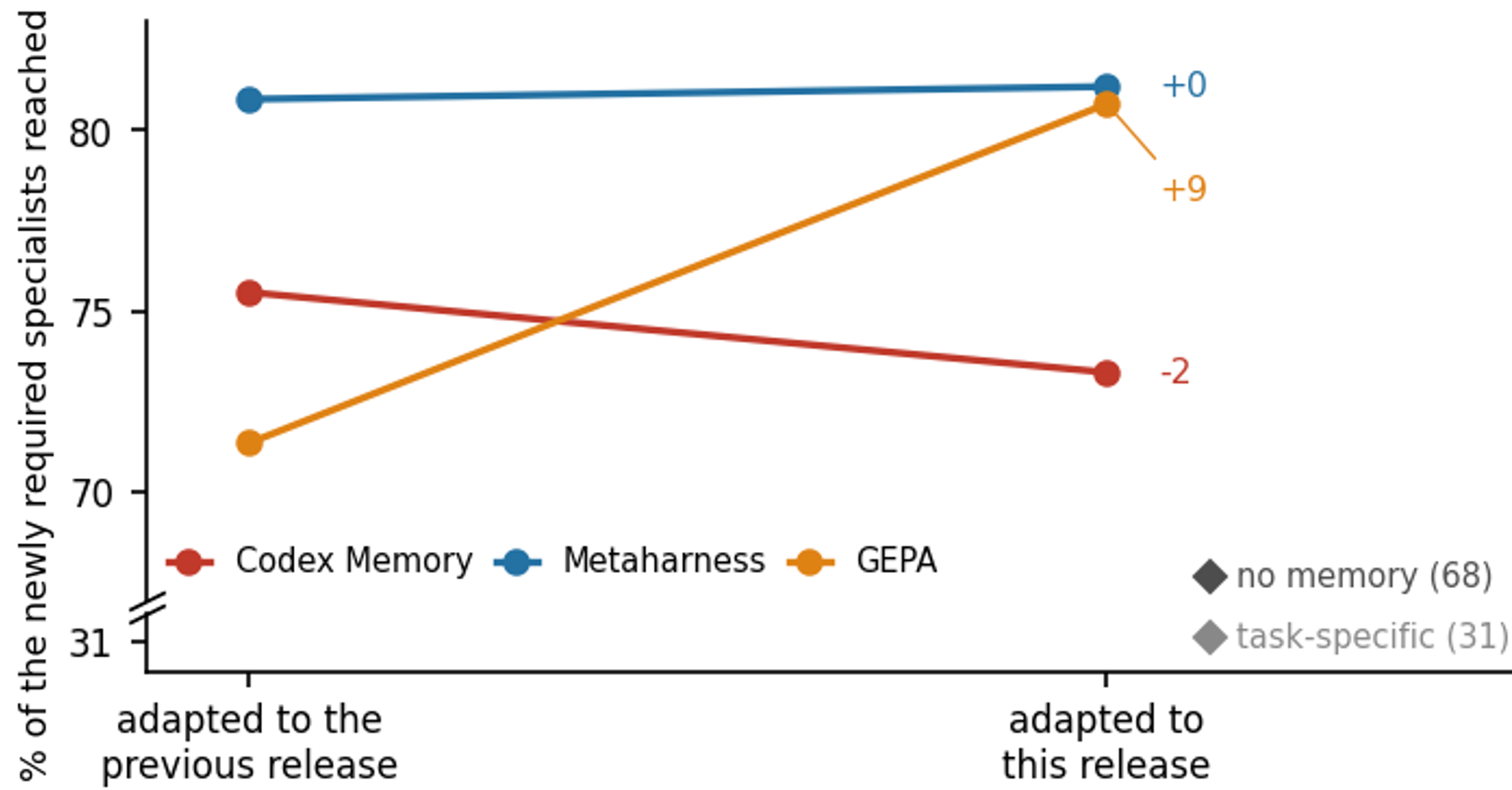}
\vspace{-.5\baselineskip}
\captionof{figure}{\small Adoption of newly introduced agents.
}
\label{fig:agent_routing_adoption}
\vspace{-1\baselineskip}
\end{figure}

\paragraph{Forward transfer arises through different mechanisms.}
We next ask whether stage-specific adaptation teaches the orchestrator to use the
specialists introduced at that stage. \Cref{fig:agent_routing_adoption} compares
new-specialist recall immediately before and after adaptation to each release.
GEPA increases recall of newly required specialists from 71\% to 81\%, showing
that its positive forward transfer is partly explained by learning to invoke
newly exposed specialists. Meta-Harness, however, already reaches approximately
81\% of the newly required specialists before adapting to the current release and
shows essentially no further improvement after adaptation. Codex Memory slightly
decreases. Positive FWT therefore does not have a single routing explanation:
GEPA improves through better adoption of newly introduced specialists, whereas
Meta-Harness's gains must arise downstream of specialist discovery, such as task
decomposition, interaction between specialists, or execution after delegation.

\paragraph{Self-evolution adaptation improves delegation completeness, not selectivity.}
As shown in \cref{fig:agent_routing_quality}, the main effect of self-evolution adaptation
is not that the lead agent becomes more precise about which agents are
relevant. Routing precision remains essentially unchanged at around 90\% from
the non-adaptive system to Codex Memory, GEPA, and Meta-Harness. Instead,
required-specialist recall increases from 79\% to 83--89\%, while exact match rises from 27\% to 33--38\% (\cref{fig:agent_routing_quality}, left).
Importantly, these gains do not come from invoking more agents. As shown in the
middle panel, GEPA reduces average delegations from 8.8 to 7.0 and repeated
delegations from 3.8 to 1.4 while increasing required-agent recall to 88\%.
The right panel further shows that higher agent coverage corresponds to higher
end-to-end success. Thus, self-evolution adaptation primarily helps the orchestrator
\emph{complete} the required delegation set and avoid redundant calls, rather
than becoming more selective about which specialists are relevant. 

\begin{figure}[h]
\centering
\vspace{-.5\baselineskip}
\includegraphics[width=0.5\columnwidth]{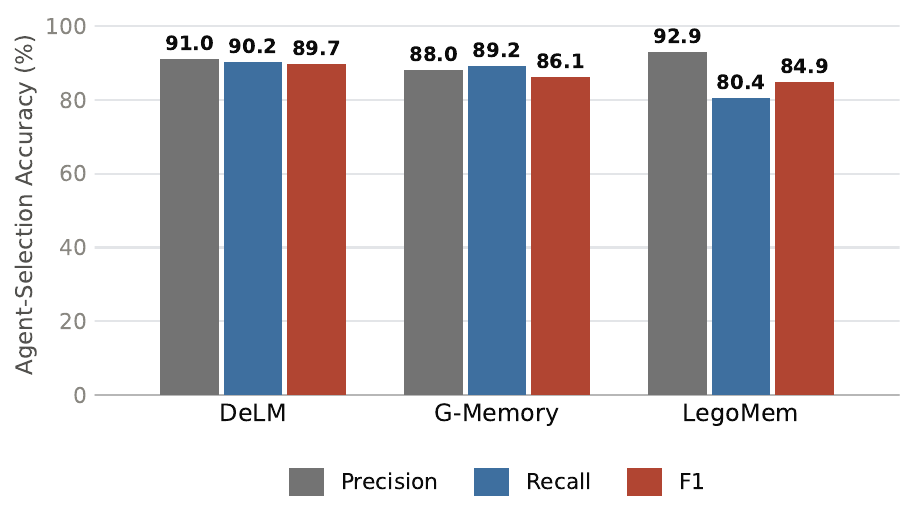}
\vspace{-.5\baselineskip}
\caption{\small Agent-selection precision, recall, and F1 by various MAS methods on the evolving-agents setting of \ourbenchmark-EOG.
}
\vspace{-0.5em}
\label{fig:mas-agent-selection}
\end{figure}

A similar pattern holds for MAS under an evolving agent pool. Across all evaluated methods, agent-selection quality is already high, with precision ranging from 88.0\% to 94.2\% and F1 from 84.9\% to 90.2\%, suggesting that identifying relevant agents is likewise not the primary bottleneck. More importantly, stronger selection accuracy does not consistently translate into better end-task performance. G-Memory, for example, has lower selection precision and F1 than DeLM, yet achieves a substantially higher final pass rate (18.2\% vs.\ 14.2\%). Conversely, LegoMem attains the highest selection precision (92.9\%) while yielding the lowest final pass rate among the three memory-based methods. Together with the SAS results, this points to a consistent conclusion: once relevant specialists can be identified with reasonably high reliability, the dominant challenge shifts from \emph{which} agents to select to \emph{how} they are used---including task decomposition, routing, invocation, and coordination of their intermediate outputs. Self-evolution therefore appears to improve orchestration mainly by making delegation more complete and effective, rather than by materially improving selection precision itself.

\paragraph{Effect of newly-added sub-agents on MAS delegation.}

We further examine whether MAS routing mechanisms generalize to \emph{newly introduced specialists} that have not appeared in earlier oracle labels, versus specialists already seen in prior versions.
\textbf{The clearest finding is not that new specialists are uniformly harder to route to, but that routing degrades as the agent pool becomes increasingly cluttered with distractors.} At the individual domain--version level, old-specialist recall exceeds new-specialist recall in most transitions; however, after pooling by task count, this pattern persists only for G-Memory (90.6\% vs.\ 86.8\%), while DeLM (87.6\% vs.\ 89.5\%) and LegoMem (72.6\% vs.\ 80.0\%) reverse.

\begin{figure}[h]
\centering
\vspace{-.5\baselineskip}
\includegraphics[width=0.5\columnwidth]{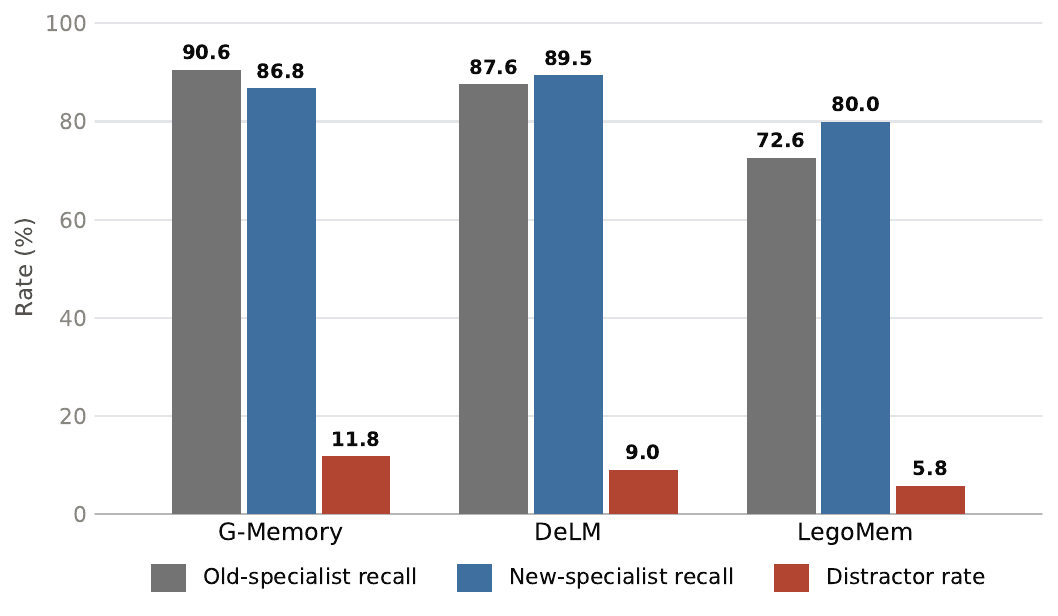}
\vspace{-.5\baselineskip}
\caption{\small Routing performance of MAS methods on old vs.\ newly introduced specialists, with distractor-selection rates.}
\vspace{-1.5em}
\label{fig:mas-new-specialist}
\end{figure}

Thus, the robust conclusion is that novel-specialist routing is locally harder in many transitions, but not consistently worse in aggregate. More importantly, the mechanisms expose a clear accuracy--caution trade-off: G-Memory attains the strongest specialist recall but also the highest distractor rate (11.8\%), indicating aggressive over-delegation; LegoMem shows the opposite behavior, with the lowest distractor rate (5.8\%) but substantially weaker recall, suggesting conservative under-delegation rather than precise routing; DeLM lies between these two extremes. \textbf{As the specialist pool grows, false delegation becomes the more consistent bottleneck:} for example, G-Memory's distractor rate on HR increases from 5.1\% at v1 to 24.7\% at v3. Overall, no mechanism simultaneously achieves both high specialist recall and low distractor routing, suggesting that the central challenge under agent evolution is not simply recognizing new specialists, but maintaining selective routing as the available agent pool expands.

\end{document}